\documentclass[a4paper,11pt]{article}
\usepackage{graphicx,floatflt,amssymb,amsmath}

\def\circa#1{\,\raise.3ex\hbox{$#1$\kern-.75em\lower1ex\hbox{$\sim$}}\,}

\newcommand{\GeV}{\,{\rm GeV}}
\newcommand{\TeV}{\,{\rm TeV}}

\newcommand{\md}[1]{\langle #1 \rangle}
\makeatletter

\def \unit{\leavevmode\hbox{\small1\kern-3.6pt\normalsize1}}
\newcommand{\beq}{\begin{equation}}
\newcommand{\eeq}{\end{equation}}
\newcommand{\bea}{\begin{eqnarray}}
\newcommand{\eea}{\end{eqnarray}}
\newcommand{\ba}{\begin{array}}
\newcommand{\ea}{\end{array}}
\newcommand{\bi}{\begin{itemize}}
\newcommand{\ei}{\end{itemize}}
\newcommand{\bn}{\begin{enumerate}}
\newcommand{\en}{\end{enumerate}}
\newcommand{\bc}{\begin{center}}
\newcommand{\ec}{\end{center}}

\newcommand{\e}{\epsilon}
\newcommand{\al}{\alpha}
\newcommand{\be}{\beta}
\newcommand{\lam}{\lambda}
\newcommand{\D}{\Delta}

\newcommand{\gsim}{\lower.7ex\hbox{$\;\stackrel{\textstyle>}{\sim}\;$}}
\newcommand{\lsim}{\lower.7ex\hbox{$\;\stackrel{\textstyle<}{\sim}\;$}}
\def \unit{\leavevmode\hbox{\small1\kern-3.6pt\normalsize1}}

\newcommand{\captionfonts}{\small}

\makeatletter  
\long\def\@makecaption#1#2{%
  \vskip\abovecaptionskip
  \sbox\@tempboxa{{\captionfonts #1: #2}}%
  \ifdim \wd\@tempboxa >\hsize
    {\captionfonts #1: #2\par}
  \else
    \hbox to\hsize{\hfil\box\@tempboxa\hfil}%
  \fi
  \vskip\belowcaptionskip}
\makeatother   

\begin{document}
\tolerance=100000
\thispagestyle{empty}
\setcounter{page}{0}

\begin{flushright}
SACLAY-T08/069\\
LPT-Orsay/08-01
\end{flushright}

\vspace*{1.5cm}

\begin{center}

{\LARGE \bf Successful Leptogenesis in $SO(10)$ Unification}\\
\vspace*{.3cm}
{\LARGE \bf with a Left-Right Symmetric Seesaw Mechanism}\\
\vspace*{2cm}

{\bf Asmaa~Abada$^{\rm a}$, Pierre~Hosteins$^{\rm b}$,
Fran\c{c}ois-Xavier~Josse-Michaux$^{\rm a}$\\
and St\'ephane~Lavignac$^{\rm c}$}\\
\vspace*{.5cm}

{\it $^{\rm a}$ Laboratoire de Physique Th\'eorique, UMR 8627, 
Universit\'e de Paris-Sud 11, B\^atiment 210,\\ 91405 Orsay Cedex,
France}\\
{\it $^{\rm b}$ Department of Physics, University of  Patras,
GR-26500 Patras, Greece}\\
{\it $^{\rm c}$ Institut de Physique Th\'eorique\! \footnote{Laboratoire
de la Direction des Sciences de la Mati\`ere du Commissariat \`a l'Energie
Atomique et Unit\'e de Recherche associ\'ee au CNRS (URA 2306).},
CEA-Saclay, F-91191 Gif-sur-Yvette Cedex, France}\\

\end{center}

\vspace*{1.5cm} 
\centerline{\bf Abstract} 
\begin{quote}
{\noindent{
We study thermal leptogenesis in a broad class of supersymmetric $SO(10)$
models with a left-right symmetric seesaw mechanism, taking into account
flavour effects and the contribution of the next-to-lightest right-handed
neutrino supermultiplet.
Assuming $M_D = M_u$ and a normal hierarchy of light
neutrino masses,
we show that  four out of the eight right-handed neutrino mass
spectra reconstructed from low-energy neutrino data can lead
to successful leptogenesis with a reheating temperature
in the ($10^9 - 10^{10}$) GeV range.
In the remaining four solutions,
leptogenesis is dominated by $N_2$ decays,
as in the type I seesaw case.
We find that some of these spectra
can generate the observed baryon asymmetry for reheating temperatures
above $10^{10}$ GeV, in contrast to the type I case.
Together with flavour effects, an accurate description of charged fermion
masses turns out to be a crucial ingredient in the analysis.}}
\end{quote}

\vfill
\eject

\newpage

\setcounter{page}{1}
\pagestyle{plain}

\section{Introduction}
\label{sec:intro}

Leptogenesis is one of the most popular mechanism
for generating the observed~\cite{WMAP} baryon asymmetry
of the universe:
\beq
  \frac{n_B - n_{\bar B}}{n_\gamma}\ =\ (6.21 \pm 0.16) \times 10^{-10}\ .
\eeq
In its simplest version~\cite{FY86}, out-of-equilibrium decays of heavy
Majorana neutrinos generate a lepton asymmetry which is
then partially converted into a baryon asymmetry by sphaleron
processes~\cite{sphalerons,KS88}. This mechanism has been
extensively studied in the last decade~\cite{DNN08}.
In particular, conditions for a successful leptogenesis have been
obtained~\cite{DI02,BDP02,GNRRS03} and many refinements
have been added, such as spectator  processes~\cite{BP01},
finite temperature corrections~\cite{GNRRS03}
and flavour effects~\cite{BCST99}.
One of these conditions is the famous Davidson-Ibarra 
bound~\cite{DI02} on the lightest right-handed neutrino mass,
$M_1 \geq {\cal O} (10^8 - 10^9)$ GeV,
which applies in the case of a hierarchical right-handed neutrino
mass spectrum. The main outcome of these studies is that thermal
leptogenesis can work with parameters consistent with the type I
seesaw~\cite{seesaw} interpretation of neutrino oscillation data.
By contrast, standard electroweak baryogenesis~\cite{KRS85} fails
to produce the observed baryon asymmetry~\cite{GHOP93,KLRS96},
and its supersymmetric version~\cite{CQW96} is successful only
in a small portion of the Minimal Supersymmetric Standard Model
(MSSM) parameter space\footnote{Electroweak baryogenesis
could however still be a viable mechanism in other extensions
of the Standard Model in which the dynamics of the electroweak
phase transition is modified (see e.g. Ref.~\cite{Cline06} for a review).}.

While thermal leptogenesis can successfully generate the baryon asymmetry
of the universe if we are free to choose the right-handed neutrino
masses and couplings (modulo the constraints coming from
neutrino masses and mixing), this might not be the case if
the seesaw mechanism is embedded into a more fundamental
theory -- typically a Grand Unified Theory (GUT) based
on the $SO(10)$ gauge group~\cite{SO10}.
In such theories, the right-handed neutrino parameters are
constrained both by the unified gauge symmetry, which implies
relations among quark and lepton mass matrices,
and by neutrino oscillation data, with no guarantee
that they fall into the range preferred by leptogenesis.
It is well known indeed~\cite {NO00} that the $SO(10)$
mass formula $M_D = M_u$
leads to a strongly hierarchical heavy neutrino mass spectrum
(except for special values of the light neutrino parameters~\cite{AFS03}),
with $M_1$ lying below the Davidson-Ibarra bound.
This conclusion can be evaded, however, if the relation $M_D = M_u$
receives large corrections from Yukawa couplings involving
a $\bf \overline{126}$ or a $\bf 120$ Higgs representation,
as in the so-called minimal $SO(10)$ model~\cite{ABMSV03}
and its extensions, or from non-renormalizable interactions~\cite{lepto_nr}.
The contribution of the next-to-lightest right-handed neutrino
to the baryon asymmetry could also, in principle, change
the above picture~\cite{DB05,Vives05}.

In this paper, we investigate another possibility
to reconcile $SO(10)$ unification with leptogenesis,
based on the left-right symmetric seesaw mechanism\footnote{A
completely different option, based on a non-standard embedding
of the Standard Model matter fields into $SO(10)$ representations,
has been explored in Refs.~\cite{Asaka03,FHLR08}. Successful
leptogenesis can also be achieved by adding chiral singlets
to supersymmetric $SO(10)$ models~\cite{lepto_nonstandard}.}.
In a broad class of $SO(10)$ models, neutrino masses
receive contributions from both the type I ~\cite{seesaw} (right-handed
neutrino exchange) and the type II~\cite{typeII} (heavy scalar $SU(2)_L$
triplet exchange) seesaw mechanisms, with both contributions
related by a left-right symmetry. As a result, for a given Dirac
mass matrix, 8 different right-handed neutrino mass spectra
are consistent with the same light neutrino mass matrix~\cite{AF05},
instead of a single one in the type I case.
It was shown in Ref.~\cite{HLS06} (albeit with a qualitative
discussion of the washout) that some of these spectra
can lead to successful leptogenesis even if the mass
relation $M_D = M_u$ holds.
This was confirmed, for the case of an inverted light neutrino
mass hierarchy, in Ref.~\cite{ABHKO06}, where the Boltzmann
equations were solved in the one-flavour approximation
(the possibility of triplet leptogenesis
has also been studied in Ref.~\cite{HKO07}).
The purpose of the present paper is to perform a more comprehensive
study of thermal leptogenesis in this class of $SO(10)$ models,
including previously missing ingredients such as
flavour effects and the contribution of the next-to-lightest
right-handed neutrino, and investigating the dependence
of the final baryon asymmetry on low- and high-energy parameters
as well as on the reheating temperature.
The necessary corrections to the GUT-scale mass relation $M_d = M_e$
are also taken into account in our analysis. Assuming $M_D = M_u$
and a normal hierarchy of light neutrino masses, we find that
successful leptogenesis is possible with a reheating temperature
in the $(10^9 - 10^{10})$ GeV range for 4 out of the 8
reconstructed right-handed neutrino spectra.
Some of the remaining 4 spectra can also generate the observed
baryon asymmetry, but for higher reheating temperatures.

The paper is organized as follows. In Section~\ref{sec:framework},
we review the left-right symmetric seesaw mechanism
in supersymmetric $SO(10)$ models, as well as
the properties of the associated  right-handed neutrino mass spectra.
In Section~\ref{sec:leptogenesis}, we write the flavour-dependent
Boltzmann equations governing leptogenesis,
including the contribution of the next-to-lightest right-handed neutrino
and of its supersymmetric partner.
In Section~\ref{sec:results}, we solve numerically the Boltzmann equations
and present our results for the final baryon asymmetry, taking into account
the corrections to the mass relation $M_d = M_e$.
In Section~\ref{sec:dependence}, we study the dependence of the final
baryon asymmetry
on the yet unmeasured light neutrino parameters,
on the high-energy Dirac couplings and on the reheating temperature.
Finally, we present our conclusions in Section~\ref{sec:conclusions}.

\section{The framework}
\label{sec:framework}

\subsection{The left-right symmetric seesaw mechanism in supersymmetric
                      $SO(10)$ models}
\label{subsec:LRseesaw}

One of the appealing features of $SO(10)$ unification is that it provides
a natural realization of the seesaw mechanism, thus yielding an elegant
explanation for the smallness of neutrino masses. Indeed, the
right-handed neutrinos needed for the (type I) seesaw mechanism
belong to the ${\bf 16}_i$ representations ($i=1,2,3$) that contain
the Standard Model matter fields, and they acquire heavy Majorana
masses at the scale where the $B-L$ symmetry (which is part of
the $SO(10)$ gauge symmetry) is broken.
Namely, the Majorana
mass matrix is generated either by the renormalizable operators
${\bf 16}_i {\bf 16}_j {\bf \overline{126}}$ or by the non-renormalizable
operators ${\bf16}_i {\bf 16}_j {\bf \overline{16}}\, {\bf \overline{16}} / \Lambda$,
in which the $SU(5)$-singlet component of the $\bf \overline{126}$
(resp. $\bf \overline{16}$) Higgs representation provides the ($B-L$)-breaking
vev. As for the Dirac neutrino mass matrix, it arises from the same $SO(10)$
operators that contribute to the charged fermion masses.
Upon integrating out the heavy Majorana neutrinos, one obtains
the well-known type I seesaw mass formula:
\beq
  M^{(I)}_\nu\ =\ - M^T_D M^{-1}_R M_D\ ,
\label{eq:Mnu_I}
\eeq
where $M_D$ and $M_R$ denote the Dirac and Majorana mass matrices,
respectively. In supersymmetric $SO(10)$ models with a $\bf \overline{126}$
Higgs representation, the light neutrino mass matrix can receive an additional
type II contribution if a $\bf 54$ Higgs representation is also present.
The role of the $\bf 54$ is to induce a coupling between the
$\bf \overline{126}$ representation, in which the $SU(2)_L$ triplet
needed for the type II seesaw mechanism lies, and a $\bf 10$ representation
containing a significant $H_u$ component\footnote{In general, $H_u$ 
is a linear combination of all $Y\! =\! + 1$ $SU(2)_L$ doublets contained
in the Higgs representations of the model.
Denoting by $H^{10}_u$ the $Y\! =\! +1$ doublet lying
in the $\bf 10$ under consideration, one can write
$H^{10}_u = \alpha_u H_u + \cdots$, where the dots stand
for heavy $Y\! =\! +1$ doublets.
We assume here that $\alpha_u \sim 1$, although strictly speaking
only $\alpha_u \neq 0$ is required.},
where $H_u$ is the MSSM Higgs doublet responsible for
up quark masses.
To see how this works, it is convenient to use a left-right symmetric
language: the $\bf \overline{126}$ contains a right-handed
triplet $\Delta^c$ with quantum numbers ${\bf (1,1,3)_{-2}}$ under
$SU(3)_C \times SU(2)_L \times SU(2)_R \times U(1)_{B-L}$,
whose vev $v_R$ is responsible for the breaking of the $B-L$ symmetry,
as well as a left-handed triplet $\Delta = {\bf (1,3,1)_{+2}}$; the $\bf 54$
contains a bitriplet $\tilde \Delta = {\bf (1,3,3)_0}$; and the ${\bf 10}$
contains a bidoublet $\Phi = {\bf (1,2,2)_0}$. The superpotential terms
relevant for the type II seesaw mechanism read:
\beq
  \frac{1}{2}\, f_{ij}\, L_i L_j \Delta
  + \frac{1}{2}\, \sigma\, \Phi \Phi \tilde \Delta
  + \tau\, \Delta \Delta^c \tilde \Delta\ ,
\label{eq:W_typeII}
\eeq
where the first term comes from the ${\bf 16}_i {\bf 16}_j {\bf \overline{126}}$
couplings, while the second and third terms come from the $\bf 10\, 10\, 54$
and $\bf 54\, \overline{126}\, \overline{126}$ couplings, respectively.
Setting $\langle (\Delta^c)^0 \rangle = v_R$ in Eq.~(\ref{eq:W_typeII})
and integrating out the heavy triplets $\Delta$ and $\tilde \Delta$,
one obtains:
\beq
  M^{(II)}_\nu\ =\ \frac{\sigma_u v^2_u}{2 M_\Delta}\, f\ ,
\label{eq:Mnu_II}
\eeq
where $\sigma_u \equiv \alpha^2_u \sigma$,
$v_u \equiv  \langle H^0_u \rangle = v \sin \beta$ ($v = 174$ GeV),
and $M_\Delta$ is an effective $SU(2)_L$ triplet mass.
The couplings of Eq.~(\ref{eq:W_typeII}) alone would give
$M_\Delta = \tau v_R$,
but the superpotential generally contains additional terms contributing
to the $SU(2)_L$ triplet mass matrix.
Depending on these,
$M_\Delta$ may be larger or smaller than $v_R$ (for $v_R \ll M_{GUT}$,
a tuning of the superpotential parameters is generally necessary
to achieve $M_\Delta < v_R$~\cite{GMN04}).
Notice that, due to the left-right symmetry embedded in the $SO(10)$
gauge symmetry, the same set of parameters
$f_{ij}$ determine the triplet couplings in Eq.~(\ref{eq:W_typeII}) and
the right-handed neutrino mass matrix, which is given by $M_R = f v_R$.
Assuming further that the Dirac neutrino mass matrix is symmetric
(which excludes a contribution from
Yukawa couplings involving a $\bf 120$ Higgs representation),
one ends up with the left-right symmetric seesaw mass formula:
\beq
  M_\nu\ =\ \frac{\sigma_u v^2_u}{2 M_\Delta}\, f
    - \frac{v^2_u}{v_R}\, Y_\nu f^{-1} Y_\nu\ ,
\label{eq:Mnu_LR}
\eeq
where we have written the Dirac mass matrix as $M_D \equiv Y_\nu v_u$.

Definite predictions for the baryon asymmetry generated via leptogenesis
require the knowledge of the masses and couplings of the heavy decaying
states. In this respect, $SO(10)$ models provide a predictive
framework, since the Dirac mass matrix is generated by the same Yukawa
couplings as the charged fermion mass matrices. In the type I seesaw case,
one can reconstruct the right-handed neutrino
mass matrix $M_R$ from the knowledge of the light neutrino mass matrix
by simply inverting Eq.~(\ref{eq:Mnu_I}), provided that the Dirac mass
matrix is known. Assuming e.g. that $M_D$ and $M_u$ only receive contribution
from renormalizable Yukawa couplings to 10-dimensional Higgs multiplets,
which implies the well-known mass relation $M_D = M_u$,
one generically obtains a strongly hierarchical right-handed neutrino mass
spectrum, with $M_1$ lying below the Davidson-Ibarra bound~\cite{DI02}
(see however Ref.~\cite{AFS03} for special situations where $M_1$ and $M_2$
can be degenerate). More generally, successful leptogenesis is difficult
to achieve in $SO(10)$ models with a type I seesaw mechanism, even
taking into account the contribution of the next-to-lightest right-handed
neutrino~\cite{HLS06} as suggested in Ref.~\cite{Vives05}.

In the left-right symmetric seesaw case, the reconstruction of the matrix $f$
that determines both the right-handed neutrino mass matrix and the triplet
couplings requires the resolution of the non-linear matrix
equation~(\ref{eq:Mnu_LR}). In Ref.~\cite{AF05}, Akhmedov and Frigerio
showed that this equation has exactly $2^n$ solutions in the
$n$ generation case, and provided explicit solutions up to $n=3$.
An alternative reconstruction procedure, which employs complex
orthogonal matrices, was proposed in Ref.~\cite{HLS06}.
There it was argued, based on a qualitative discussion of the washout,
that this multiplicity of solutions
makes it  possible for leptogenesis to be successful in $SO(10)$ models
with a left-right symmetric seesaw mechanism. The purpose of the present
paper is to put this statement on a quantitative basis, and to prove in particular
that successful leptogenesis is indeed possible for ``mixed'' solutions
in which neither the type I not the type II seesaw contribution dominates
in the light neutrino mass matrix.

\subsection{Reconstruction procedure}
\label{subsec:procedure}

Before presenting our study, let us briefly recall the reconstruction procedure
of Ref.~\cite{HLS06}.
Our starting point is the left-right symmetric seesaw formula~(\ref{eq:Mnu_LR}),
in which both $f$ and $Y_\nu$ are complex symmetric matrices.
We want to reconstruct $f$ for a given pattern of light neutrino
masses and lepton mixing, assuming that $Y_\nu$
is known in a basis in which the charged lepton mass matrix is diagonal.
For concreteness, we work in the $3$-family case, but the procedure
applies to any number of neutrino families.

In order to solve Eq.~(\ref{eq:Mnu_LR}), we first rewrite it as
\begin{equation}
  Z\ =\ \alpha X - \beta X^{-1}\, ,
\label{eq:master_equation}
\end{equation}
with $\alpha \equiv \sigma_u v^2_u / (2 M_\Delta)$, $\beta \equiv v^2_u / v_R$
and
\begin{equation}
  Z\ \equiv\ N_\nu^{-1} M_\nu (N_\nu^{-1})^T\, , \quad
  X\ \equiv\ N_\nu^{-1} f (N_\nu^{-1})^T\, ,
\label{eq:def_Z_X}  
\end{equation}
where $N_\nu$ is a matrix such that $Y_\nu = N_\nu N_\nu^T$,
and $Y_\nu$ is assumed to be invertible.
Being complex and symmetric, $Z$ can be diagonalized by a complex
orthogonal matrix if its eigenvalues (i.e. the roots of the
characteristic polynomial $\det (Z - z \mathbf{1}) = 0$) are all distinct:
\begin{equation}
 Z\ =\ O_Z\, \mbox{Diag}\, (z_1, z_2, z_3)\, O^T_Z\ ,  \qquad  |z_1| < |z_2| < |z_3|\ ,
   \qquad  O_Z O^T_Z\ =\ \mathbf{1}\ .
\label{eq:diag_Z} 
\end{equation}
Then Eq. (\ref{eq:master_equation}) can be solved for $X$ in a straightforward
manner, by noting that $X$ is diagonalized by the same complex orthogonal
matrix as $Z$. Upon an $O_Z$ transformation, Eq. (\ref{eq:master_equation})
reduces to 3 independent quadratic equations for the eigenvalues of $X$:
\beq
  z_i\ =\ \alpha x_i - \beta x^{-1}_i\ .
\label{eq:}
\eeq
For a given choice of ($x_1$, $x_2$, $x_3$),
the solution of Eq.~(\ref{eq:Mnu_LR}) is given by:
\begin{equation}
  f\ =\ N_\nu\, O_Z\, \mbox{Diag}\, (x_1, x_2, x_3)\, O^T_Z\, N_\nu^T\ .
\label{eq:f}
\end{equation}
The right-handed neutrino masses $M_i = f_i v_R$ are obtained
by diagonalizing $f$ with a unitary matrix:
\beq
  f\ =\ U_f\, \mbox{Diag}\, (f_1, f_2, f_3)\, U^T_f\ ,  \qquad  f_1 < f_2 < f_3\ ,
    \qquad  U_f U^\dagger _f = \mathbf{1}\ ,
\label{eq:diag_f}
\eeq
where the $f_i$ are chosen to be real and positive.
The matrix $U_f$ relates the original basis for right-handed neutrinos,
in which $Y_\nu$ is symmetric, to their mass eigenstate basis. It can be
used to express the Dirac couplings in terms of charged lepton
and right-handed neutrino mass eigenstates, as
$\lambda \equiv U^\dagger_f Y_\nu$.

Since each equation $z_i = \alpha x_i - \beta x^{-1}_i$ has
two solutions $x^-_i$ and $x^+_i$, there are $2^3 = 8$ different
solutions for the matrix $f$, which we label in the following way:
$(+,+,+)$ refers to the solution $(x^+_1, x^+_2, x^+_3)$, $(+,+,-)$
to the solution $(x^+_1, x^+_2, x^-_3)$, and so on.
It is convenient to define $x^-_i$ and $x^+_i$ such that, in the
$4 \alpha \beta \ll |z_i|^2$ limit:
\begin{equation}
  x^-_i\ \simeq\ - \frac{\beta}{z_i}\ ,  \qquad  x^+_i\ \simeq\ \frac{z_i}{\alpha}\ .
\label{eq:x_pm_limit}
\end{equation}
We will refer to $x^-_i$ as the ``type I branch'' and to $x^+_i$
as the ``type II branch''. This terminology is motivated by the fact
that solutions $(-,-,-)$ and $(+,+,+)$ reduce to the ``pure'' type I
and type II cases in the large $v_R$ limit (defined by $4 \alpha \beta \ll |z_1|^2$,
or equivalently $v_R \gg 2 \sigma_u v^4_u / M_\Delta |z_1|^2$):
\begin{eqnarray}
  f^{(-,-,-)} & \stackrel{4 \alpha \beta \ll |z_1|^2}{\longrightarrow} &
  -\, \frac{v^2_u}{v_R}\, Y_\nu M^{-1}_\nu Y_\nu\ ,  \\
  f^{(+,+,+)} & \stackrel{4 \alpha \beta \ll |z_1|^2}{\longrightarrow} &
  \frac{M_\nu}{\alpha}\ .
\end{eqnarray}
The remaining 6 solutions correspond to mixed cases where
the light neutrino mass matrix receives significant contributions
from both types of seesaw mechanisms.
In the opposite, small $v_R$ limit ($|z_3|^2 \ll 4 \alpha \beta$), one has
$x^\pm_i\ \simeq\ \pm\, \mbox{sign} (\mbox{Re} (z_i)) \sqrt{\beta / \alpha}$,
which indicates a partial cancellation between the type I and type II
contributions to light neutrino masses.
Finally, in the region of intermediate $v_R$ values,
$|z_1|^2 < 4 \alpha \beta < |z_3|^2$,
both the type I and the type II seesaw mechanisms give
significant contributions to the light neutrino mass matrix.

\subsection{Properties of the reconstructed right-handed neutrino spectra}
\label{subsec:HLS}

The above procedure can be used to determine the a priori unknown
$f_{ij}$ couplings in theories which predict the Dirac matrix $Y_\nu$,
taking low-energy neutrino data as an input. 
In Ref.~\cite{HLS06}, it was applied to supersymmetric $SO(10)$ models
with two $\bf 10$'s, a $\bf 54$ and a pair of $\bf  126 \oplus \overline{126}$
representations in the Higgs sector,
but no $\bf 120$ representation (as required by the left-right symmetric
seesaw mechanism).
In this subsection, we recall the main properties of the reconstructed
right-handed neutrino mass spectra.

Given the above assumptions, the most general
renormalizable Yukawa couplings read:
\beq
  Y^{(1)}_{ij}\, {\bf 16}_i {\bf 16}_j {\bf 10}_1 + Y^{(2)}_{ij}\, {\bf 16}_i {\bf 16}_j {\bf 10}_2
  + f_{ij}\, {\bf 16}_i {\bf 16}_j {\bf \overline{126}}\ ,
\label{eq:Yukawa_SO10}
\eeq
where $Y^{(1)}$, $Y^{(2)}$ and $f$ are complex symmetric matrices.
Assuming that the $SU(2)_L$ doublet components of the $\bf \overline{126}$
do not acquire a vev,
Eq.~(\ref{eq:Yukawa_SO10}) leads to the following GUT-scale
mass relations:
\beq
  M_u\ =\ M_D\ , \qquad M_d\ =\ M_e\ .
\label{eq:M_10}
\eeq
It is well known that the second relation is in conflict with experimental data
and needs to be corrected.
In the absence of a $\bf 210$ Higgs representation that would induce
vev's for the doublet components of the $\bf \overline{126}$~\cite{BM92},
this must be done by non-renormalizable interactions. We postpone
the discussion of this issue to Section~\ref{sec:results}, and assume
for the time being that Eq.~(\ref{eq:M_10}) holds.

The inputs in the reconstruction procedure are the matrices $Y_\nu$
and $M_\nu$ at the seesaw scale.
Neglecting the running of $Y_\nu$ between the GUT scale and the
seesaw scale, Eq.~(\ref{eq:M_10}) yields, in the basis
for the $\bf 16$ matter representations in which $M_e$
is diagonal with real positive entries:
\beq
  Y_\nu\ =\ U^T_q \hat Y_u U_q\ , \qquad U_q\ =\ P_u V_{CKM} P_d\ ,
  \qquad \hat Y_u\ =\ \mbox{Diag}\, (y_u, y_c, y_t)\ ,
\label{eq:Y_nu}
\eeq
where $V_{CKM}$ is the CKM matrix and $y_{u, c, t}$ are the up quark
Yukawa couplings, all renormalized at the GUT scale.
The presence of two diagonal matrices of phases $P_u$ and $P_d$
in Eq.~(\ref{eq:Y_nu}) is due to the fact that the $SO(10)$ symmetry prevents
independent rephasing of right-handed and left-handed quark fields.
In the same basis, the light neutrino mass matrix generated
from the seesaw mechanism reads:
\beq
  M_\nu\ =\ U^\star_l \hat M_\nu U^\dagger_l\ , \qquad U_l\ =\ P_e V_{PMNS} P_\nu\ ,
  \qquad \hat M_\nu\ =\ \mbox{Diag}\, (m_1, m_2, m_3)\ ,
\label{eq:M_nu}
\eeq
where $V_{PMNS}$ contains a single, Dirac-type phase $\delta_{PMNS}$,
and $P_e$ and $P_\nu$ are two diagonal matrices of phases. With the
convention that $P_\nu$ contains only two phases,
$U_{PMNS} \equiv V_{PMNS} P_\nu$ is the PMNS lepton
mixing matrix and $m_{1,2,3}$
are the light neutrino masses, all renormalized at the seesaw
scale\footnote{Strictly speaking, Eq.~(\ref{eq:Mnu_LR}) involves the decoupling
of four states at scales which can differ by several orders of magnitude.
We neglect the associated radiative corrections here and, for simplicity,
identify the seesaw scale with the GUT scale in the following.}.
The two phases in $P_\nu$ are the physical CP-violating phases
associated with the Majorana nature of the light neutrinos,
while the three phases contained in $P_e$, analogous to the five independent
phases contained in $P_u$ and $P_d$, are pure high-energy phases.
Once the input values for $(y_u, y_c, y_t)$, $(m_1, m_2, m_3)$, $V_{CKM}$
and $U_{PMNS}$ at the GUT scale are fixed, the 8 different $f$ matrices
can be reconstructed as a function of $\alpha$, $\beta$
(or, equivalently, of the $B-L$ breaking scale $v_R$ and $\beta / \alpha$)
and of the high-energy phases contained in $P_u$, $P_d$ and $P_e$.
Notice that, as long as Eq.~(\ref{eq:M_10}) holds, the reconstructed $f_{ij}$
couplings depend on the combination of phases $P_d P_e$ rather than
on $P_d$ and $P_e$ separately; hence the number of independent
high-energy phases reduces to five.

Let us specify the input values that we are going to use in this paper.
For the quark masses and the CKM parameters at the $M_Z$ scale,
we take the central values given in Refs.~\cite{Jamin06} and~\cite{PDG06},
respectively:
\bea
&  m_u\, (M_Z)\, =\, 1.7\, \mbox{MeV}\, , \quad
  m_c\, (M_Z)\, =\, 0.62\, \mbox{GeV}\, , \quad
  m_t\, (M_Z)\, =\, 171\, \mbox{GeV}\, ,  \\
&  m_d\, (M_Z)\, =\, 3.0\, \mbox{MeV}\, , \quad
  m_s\, (M_Z)\, =\, 54\, \mbox{MeV}\, , \quad
  m_b\, (M_Z)\, =\, 2.87\, \mbox{GeV}\, ,  \\
&  A\, (M_Z)\, =\, 0.818\, ,  \quad  \lambda\, =\, 0.2272\, ,  \quad
  \bar \rho\, =\, 0.221\, ,  \quad  \bar \sigma\, =\, 0.340\, .
\eea
These values, together with the inputs for the lepton mass and mixing parameters,
are subsequently evolved to the GUT scale $M_{GUT} \simeq 2 \times 10^{16}$ GeV
using the Mathematica package REAP~\cite{REAP}
with an effective supersymmetric threshold $M_{SUSY} = 1$ TeV and $\tan \beta = 10$.
Our reference light neutrino spectrum is a normal hierarchical
spectrum with $m_1 = 10^{-3}$ eV and $\theta_{13} = 0$.
For the observed oscillation parameters, we take the best fit values
of Ref.~\cite{FLMP05}:
\bea
  & \Delta m^2_{32}\, \equiv\, m^2_3 - m^2_2\, =\, 2.4 \times 10^{-3}\, \mbox{eV}^2 ,
  \quad  \sin^2 \theta_{23}\, =\, 0.44\, , &
  \label{eq:atmo_fit}  \\
  & \Delta m^2_{21}\, \equiv\, m^2_2 - m^2_1\, =\, 7.92 \times 10^{-5}\, \mbox{eV}^2 ,
  \quad \sin^2 \theta_{12}\, =\, 0.314\, . &
\label{eq:solar_fit}
\eea
The Dirac-type phase $\delta_{PMNS}$ and the two relative Majorana phases
contained in $P_\nu$ are treated as free parameters, as well as
the high-energy phases contained in $P_u$, $P_d$ and $P_e$.
We denote these phases by  $\Phi^{u,d,\nu,e}_i$, $i =1,2,3$ (some
of which are redundant), with
$P_u \equiv \mbox{Diag}\, (e^{i \Phi^u_1}, e^{i \Phi^u_2}, e^{i \Phi^u_3})$,
$P_d \equiv \mbox{Diag}\, (e^{i \Phi^d_1}, e^{i \Phi^d_2}, e^{i \Phi^d_3})$,
and so on.

\begin{figure}[h!]
\begin{center}
\includegraphics[scale=0.5]{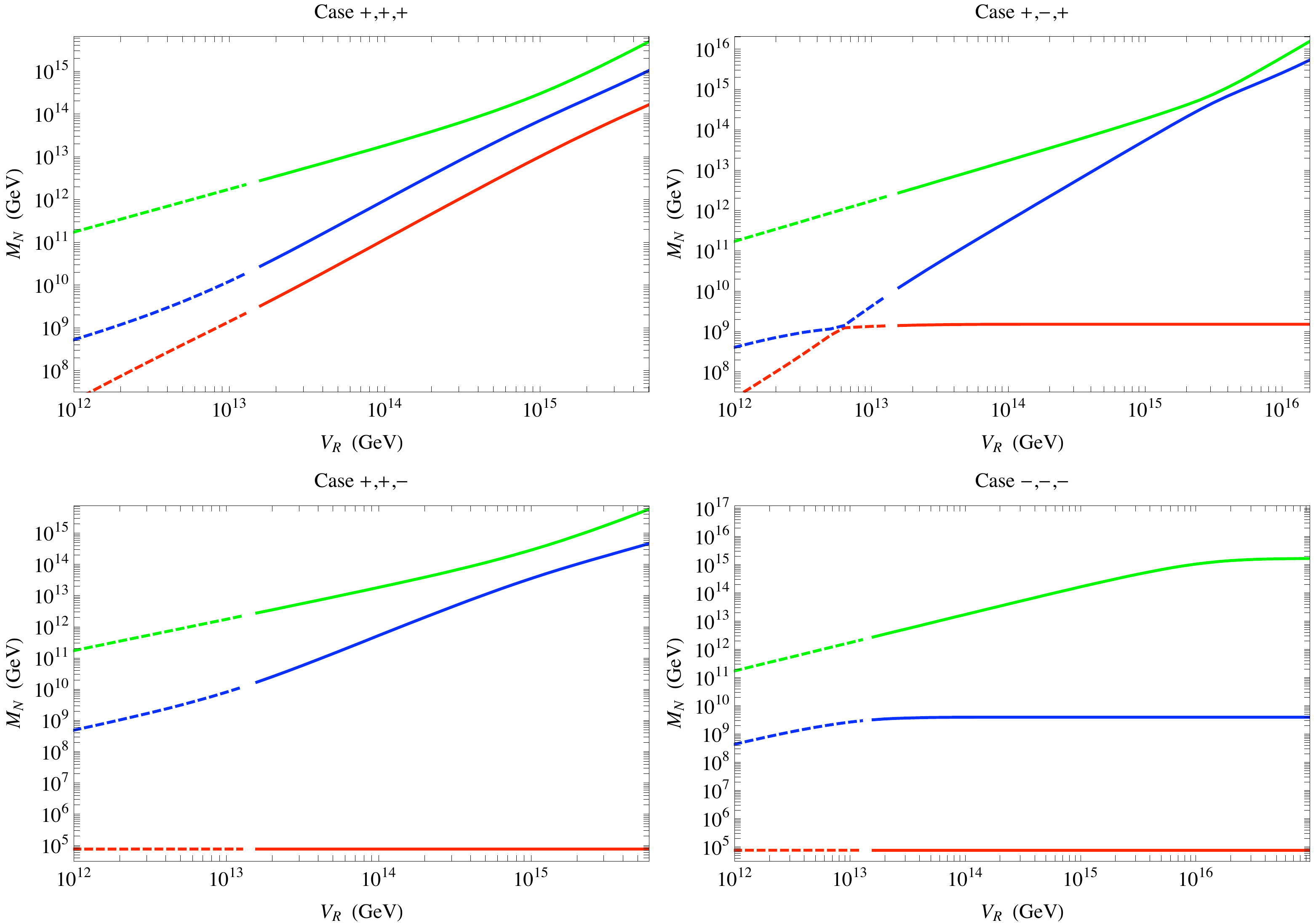}
\caption{Right-handed neutrino masses as a function of $v_R$
in solutions $(+,+,+)$, $(+,-,+)$, $(+,-,-)$ and $(-,-,-)$. Inputs: hierarchical light
neutrino masses with $m_1 = 10^{-3}$ eV, oscillation parameters as specified
in the text, and no CP violation besides the CKM phase ($\delta_{PMNS} = \Phi^u_i
= \Phi^d_i = \Phi^\nu_i = \Phi^e_i = 0$); $\beta / \alpha = 0.1$. The range
of variation of $v_R$ is restricted from above by the requirement that
$|f_{ij}| \leq 1$. Dashed lines indicate a cancellation at a stronger level than 1\%
between the type I and type II contributions to the light neutrino mass matrix.}
\label{fig:spectra}
\end{center}
\end{figure}

Fig.~\ref{fig:spectra} shows four out of the eight right-handed neutrino
spectra reconstructed from the above inputs as a function of $v_R$,
assuming $\beta / \alpha = 0.1$.
The 8 different solutions can be distinguished by the behaviour
of each $M_i$ ($i=1,2,3$) as a function of $v_R$.
The 4 solutions with $x_3 = x^-_3$ are
characterized by a constant value of the lightest right-handed
neutrino mass, $M_1 \approx 7 \times 10^4$ GeV; among them,
the 2 solutions with $x_2 = x^-_2$ also have
$M_2 \approx 4 \times 10^9$ GeV, while the 2 solutions
with $x_2 = x^+_2$ have a rising $M_2$.
It is interesting to note that $7 \times 10^4$ GeV and
$4 \times 10^9$ GeV are nothing but the type I values
of $M_1$ and $M_2$, respectively. This is actually not a coincidence,
but a direct consequence of Eq.~(\ref{eq:x_pm_limit}): besides
the fact that solution $(-,-,-)$ reduces to the type I case in the
large $v_R$ limit, some of the properties of the type I right-handed
neutrino mass spectrum are inherited by the four solutions\footnote{A
comment about our notation might be necessary: the prime symbol
in $\pm'$ means that the second sign in $(\pm,\pm',-)$ is not correlated
with the first one, i.e. $(\pm,\pm',-)$ refers to the four solutions
$(+,+,-)$, $(+,-,-)$, $(-,+,-)$ and $(-,-,-)$.} $(\pm,\pm',-)$.
The 2 solutions with $x_3 = x^+_3$ and $x_2 = x^-_2$, on the other hand,
are characterized by $M_1 \approx 2 \times 10^{9}$ GeV, and
the 2 solutions with $x_3 = x^+_3$ and $x_2 = x^+_2$ by a rising $M_1$.
In the large $v_R$ region, solution $(+,+,+)$ approaches
the type II case, as can be seen from the fact that
$M_1 : M_2 : M_3\ (\propto f_1 : f_2 : f_3) \propto m_1 : m_2 : m_3$.
It should be stressed that the choice of $\beta / \alpha$ does not affect
the shape of the curves $M_i = M_i (v_R)$, but only their position
along the horizontal axis. For instance, setting $\beta / \alpha = 1$
instead of $0.1$ would shift the curves in Fig.~\ref{fig:spectra} according
to $v_R \rightarrow \sqrt{0.1}\, v_R$.
This is due to the fact that, while $f$ depends on $\alpha$ and $\beta$
separately, $M_R = f v_R$ only depends on the combination $\alpha \beta$.
In the following we choose $\beta / \alpha = 0.1$ for numerical convenience,
but it should be clear that our results will not depend on this choice.

The implications for leptogenesis of these solutions were discussed
in Ref.~\cite{HLS06}. Since $M_\Delta = (\beta / \alpha)\, \sigma_u v_R / 2
\lesssim (\beta / \alpha)\, v_R$ and all solutions satisfy $M_1 \ll v_R$,
one can safely assume that the $SU(2)_L$ triplet
is heavier than the lightest right-handed neutrino. Then
the dominant contribution to the lepton asymmetry comes from
the out-of-equilibrium decays of $N_1$ and $\tilde N_1$, except
in some cases where the contribution of $N_2$ and $\tilde N_2$
actually dominates.
One can distinguish between three different behaviours
(from now on, $N_i$ will refer both to the $i$th right-handed neutrino
and to its supersymmetric partner):
\begin{itemize}
\item {\bf solutions $(\pm,\pm',-)$:} the four solutions characterized by
a low value of $M_1$ fail to generate the observed baryon asymmetry
from $N_1$ decays. In these solutions, the CP asymmetry
in $N_1$ decays always lies below ${\cal O} (10^{-10})$,
irrespective of the choice of the high-energy phases.
This situation is similar to the one encountered in the type I case.
In principle, $N_2$ decays could generate a large asymmetry
in a lepton flavour that is only mildly washed out by $N_1$ decays
and inverse decays, thus leading to successful leptogenesis~\cite{Vives05}.
Estimates of this effect tend to show, however, that it
is unlikely to work in solution $(-,-,-)$.
\item {\bf solutions  $(\pm,+,+)$:} in the two solutions characterized
by a rising $M_1$, the CP asymmetry in $N_1$ decays grows with
$v_R$. Successful leptogenesis then becomes possible for large
values of $v_R$, i.e. in the region where the type II seesaw
contribution dominates in the light neutrino mass matrix. 
However, one finds a tension between successful leptogenesis,
which requires $M_1 \gtrsim 10^{10}$ GeV, and
the gravitino  overproduction problem~\cite{gravitino_problem},
which imposes an upper bound $T_{RH} \lesssim (10^9 - 10^{10})$ GeV
on the reheating temperature~\cite{thermal_production}.
\item {\bf solutions $(\pm,-,+)$:} the two solutions characterized
by $M_1\sim10^9$ GeV can lead to a relatively large $CP$ asymmetry
in $N_1$ decays
without conflicting with the gravitino constraint.
However, the washout of the generated lepton
asymmetry by lepton number violating processes tends to be large.
To determine whether the observed baryon asymmetry can indeed
be generated, one must integrate numerically the Boltzmann equations.
\end{itemize}
A general feature of all solutions is that lepton number violating
processes tend to efficiently wash out the generated lepton asymmetry.
This can be traced back to the relation $M_D = M_u$, which implies
that at least one of the Dirac couplings is of the order of the top quark
Yukawa coupling.
As a consequence,
predictions for leptogenesis depend on the details of the
dynamics encoded in the Boltzmann equations.

It is clear from the above discussion that, for most solutions, the qualitative
analysis of Ref.~\cite{HLS06} is not sufficient to tell whether leptogenesis
can indeed be successful. The purpose of the present paper is to
perform a careful, quantitative study of leptogenesis in supersymmetric
$SO(10)$ models with a left-right symmetric seesaw mechanism,
taking into account the lepton flavour dynamics~\cite{EMX03}--\cite{Antusch07},
as well as the contribution of the next-to-lightest right-handed
neutrino supermultiplet~\cite{DB05,Vives05,SY07,BdB0603,EGNN06}.
As is well known from studies performed in the type I case,
flavour effects can significantly enhance the final baryon asymmetry
if there is a hierarchy between the washout parameters for different
lepton flavours~\cite{issues,NNRR06,matters}, and their impact might
be crucial for solutions $(\pm,\pm',-)$ and $(\pm,-,+)$.
Furthermore, the contribution of the next-to-lightest right-handed
(s)neutrino can be relevant both for $M_1 \approx M_2$ and for
$M_1 \ll M_2$, provided in the latter case that $N_1$-related
washout effects are weak.
The Boltzmann equations including all these ingredients will be
presented in the next section.
Finally, the corrections to the GUT-scale mass relation $M_d = M_e$
needed to account for the measured down quark and charged
lepton masses will modify the reconstructed seesaw parameters
and affect the final baryon asymmetry. These effects will be taken
into account in Section~\ref{sec:results}.

\section{Boltzmann equations}
\label{sec:leptogenesis}

In this section, we write the Boltzmann equations that govern thermal
leptogenesis in the class of supersymmetric $SO(10)$ models described above. 
The relevant heavy degrees of freedom are the three right-handed neutrinos
$N_{1,2,3}$ and the scalar Higgs triplet $\Delta$, as well as their
supersymmetric partners. In principle, all these states could contribute
to the generation of the lepton asymmetry through their decays. However,
due to the strong hierarchy among their masses, only the decays
of $N_1$ and $N_2$ and of their scalar partners are relevant in
practice\footnote{Indeed, all solutions satisfy $M_{1, 2} \ll M_3, M_\Delta$
over a large range of values for $v_R$ (assuming
$M_\Delta \sim (\beta/\alpha)\, v_R$ and $\beta / \alpha$ not too small),
hence the lepton asymmetry generated
in $\Delta$ and $N_3$ decays is washed out by $N_2$- and $N_1$-related
processes. In the large $v_R$ region of some solutions, one has
instead $M_2 \sim M_3$ ($M_\Delta$), but $N_2$ and $N_3$ ($\Delta$)
turn out to be too heavy to be thermally produced after reheating.
Indeed, one typically has $M_2 > 10^{12}$ GeV in this case, 
while the upper bound on the reheating temperature associated with
the gravitino overproduction problem is ${\cal O} (10^9 - 10^{10})$ GeV.}.

Let us first consider the CP asymmetries in heavy (s)neutrino decays.
The relevant superpotential terms, written in the basis
of charged lepton and right-handed neutrino mass eigenstates, read:
\beq
  W_{seesaw}\ =\ \lambda_{i\alpha} N^c_i L^T_\alpha i \sigma^2 H_u
    + \frac 1 2 M_i N^c_i N^c_i
    + \frac 1 2 f_{\alpha \beta} L^T_\alpha i \sigma^2 \Delta L_\beta
    + \frac 1 2 \sigma_u H^T_u i \sigma^2 \bar \Delta H_u
    + M_\Delta \mbox{Tr} (\Delta \bar \Delta)\ ,
\eeq
where $N^c_i$, $L_\alpha$ and $H_u$ are the right-handed neutrino,
lepton doublet and Higgs doublet superfields, respectively, and
\beq
  \Delta \equiv \frac{\vec \sigma}{\sqrt{2}} \cdot \vec{\Delta} =
    \left( \ba{cc}  \Delta^+ / \sqrt{2} & \Delta^{++}  \\
    \Delta^0 & - \Delta^+ / \sqrt{2}  \ea \right) , \qquad
  \bar \Delta \equiv \frac{\vec \sigma}{\sqrt{2}} \cdot \vec{\bar \Delta} =
    \left( \begin{array}{cc}  \bar \Delta^- / \sqrt{2} & \bar \Delta^0  \\
    \bar \Delta^{--} & - \bar \Delta^- / \sqrt{2}  \end{array} \right) .
\eeq
Being Majorana particles, the heavy neutrinos can decay both into
$\ell_\alpha H_u$ (lepton + Higgs boson), $\tilde \ell_\alpha \tilde H_u$
(slepton + higgsino) and into the CP-conjugated final states
$\bar \ell_\alpha H^*_u$ and $\tilde \ell^*_\alpha \bar{\tilde H}_u$.
Their scalar partners $\widetilde{N^c_i}$ are not CP eigenstates and
have only 2 two-body decay modes,
$\widetilde{N^c_i} \rightarrow \bar \ell_\alpha \bar {\tilde H}_u$ and
$\widetilde{N^c_i} \rightarrow \tilde \ell_\alpha H_u$.
The corresponding tree-level decay rates are~\cite{CRV96}:
\beq
  \Gamma (N_i \rightarrow \ell_\alpha H_u)\,
    =\, \Gamma (N_i \rightarrow \bar \ell_\alpha H^*_u)\,
    =\, \Gamma (N_i \rightarrow \tilde \ell_\alpha \tilde H_u)\,
    =\, \Gamma (N_i \rightarrow \tilde \ell^*_\alpha \bar{\tilde H}_u)\,
    =\, \frac{M_i}{16 \pi}\, |\lambda_{i \alpha}|^2\ ,
\eeq
\beq
  \Gamma (\widetilde{N^c_i} \rightarrow \bar \ell_\alpha \bar {\tilde H}_u)\,
    =\,  \Gamma (\widetilde{N^c_i} \rightarrow \tilde \ell_\alpha H_u)\,
    =\, \frac{M_i}{8 \pi}\, |\lambda_{i \alpha}|^2\ .
\eeq
Supersymmetry implies the equality of the total decay widths,
$\Gamma_{N_i} = \Gamma_{\widetilde{N^c_i}}
= M_i (\lambda \lambda^\dagger)_{ii} / 4 \pi$.
At the one-loop level, an asymmetry between CP-conjugated
decay channels arises from the interference of tree-level
and one-loop Feynman diagrams.
One can define 4 different types of flavour-dependent CP asymmetries
which, due to supersymmetry, are given by the same quantities
$\epsilon_{i \alpha}$:
\bea
  \epsilon_{i \alpha} & \equiv & \frac{\Gamma (N_i \rightarrow \ell_\alpha H_u)
    - \Gamma (N_i \rightarrow \bar \ell_\alpha H^*_u)}
    {\Gamma (N_i \rightarrow \ell H_u)
    + \Gamma (N_i \rightarrow \bar \ell H^*_u)}\
  =\ \frac{\Gamma (N_i \rightarrow \tilde \ell_\alpha \tilde H_u)
    - \Gamma (N_i \rightarrow \tilde \ell^*_\alpha \bar{\tilde H}_u)}
    {\Gamma (N_i \rightarrow \tilde \ell \tilde H_u)
    + \Gamma (N_i \rightarrow \tilde \ell^* \bar{\tilde H}_u)}  \nonumber  \\
  & = & \frac{ \Gamma (\widetilde{N^c_i} \rightarrow \bar \ell_\alpha \bar {\tilde H}_u)
    - \Gamma (\widetilde{N^c_i}^* \rightarrow \ell_\alpha \tilde H_u)}
    {\Gamma (\widetilde{N^c_i} \rightarrow \bar \ell \bar {\tilde H}_u)
    + \Gamma (\widetilde{N^c_i}^* \rightarrow \ell \tilde H_u)}\
  =\ \frac{\Gamma (\widetilde{N^c_i} \rightarrow \tilde \ell_\alpha H_u)
    - \Gamma (\widetilde{N^c_i}^* \rightarrow \tilde \ell^*_\alpha H^*_u)}
    {\Gamma (\widetilde{N^c_i} \rightarrow \tilde \ell H_u)
    + \Gamma (\widetilde{N^c_i}^* \rightarrow \tilde \ell^* H^*_u)}\ ,
\eea
where $\Gamma (N_i \rightarrow \ell H_u) \equiv
\sum_\beta \Gamma (N_i \rightarrow \ell_\beta H_u)$, etc.
The $\epsilon_{i \alpha}$'s receive both a type I contribution
(right-handed neutrino-induced vertex and self-energy corrections)
and a type II contribution (triplet-induced vertex correction):
\bea
\e_{i\al}\ =\ \e_{i\al}^{I}+\e_{i\al}^{II}\ .
\eea
The type I contribution reads~\cite{CRV96}:
\bea
\label{ECPI}
\e_{i\al}^{I}\ =\ \frac{1}{8 \pi} \sum_{j\neq i}\, \frac{\mbox{Im}[\lam_{i \al} (\lam \lam^{\dagger})_{i j} \lam^*_{j \al}]}{(\lam \lam^{\dagger})_{i i}}\, f_{I}\! \left(\frac{M_{j}^{2}}{M_{i}^{2}}\right) ,
\label{eq:epsilon_I}
\eea
while the type II contribution is given by~\cite{HS03,AK04}:
\bea
\e_{i\al}^{II}\ =\ \frac{3}{8 \pi}  \sum_\beta\, \frac{\mbox{Im}[\lam_{i \beta} (M^{II}_\nu)^*_{\beta \al} \lam_{i \al}]}{(\lam \lam^{\dagger})_{i i}}\, \frac{M_{i}}{v^2_{u}}\, f_{II}\! \left(\frac{M_{\D}^{2}}{M_{i}^{2}}\right) .
\label{eq:epsilon_II}
\eea
In Eqs.~(\ref{eq:epsilon_I}) and~(\ref{eq:epsilon_II}), the loop functions
are given by:
\beq
f_{I}(x)\ =\ \sqrt{x}\, \left(\frac{2}{1-x}-\ln{\left(\frac{1+x}{x}\right)}\right)\ \stackrel{x\gg 1}{\longrightarrow}\ -\frac{3}{\sqrt{x}}  \ ,
\label{eq:f_I}
\eeq
\beq
f_{II}(y)\ =\ y\, \ln{\left(\frac{1+y}{y}\right)}\ \stackrel{y\gg 1}{\longrightarrow}\ 1 \ .
\label{eq:f_II}
\eeq
Eqs.~(\ref{eq:epsilon_I}) and~(\ref{eq:f_I}) are
valid for hierarchical right-handed neutrinos; in the case of a partial degeneracy,
$M_{1} \simeq M_{2} \ll M_{3}$, they must be modified. Following Ref.~\cite{Pilaftsis},
we take:
\bea
\e_{i \al}^{I}\ =\ \frac{1}{8 \pi} \frac{1}{(\lam \lam^{\dagger})_{i i} } \sum_{j\neq i}\, \mbox{Im} \left( \lam_{i \al}\lam^{*}_{j \al} \left\lbrace(\lam \lam^{\dagger})_{i j} ( C_{v}^{i j}+2 C_{s,a}^{i j}) +(\lam \lam^{\dagger})_{j i} 2 C_{s,b}^{i j} \right\rbrace \right) ,
\label{eq:epsilon_I_deg}
\eea
where $C_v^{i j}$, $C_{s,a}^{i j}$ and $C_{s,b}^{i j}$ are functions of
$M^2_j / M^2_i$. $C_v^{i j}$ is the usual vertex function:
\bea
C_{v}^{i j} (x)\ =\ -\sqrt{x}\, \ln \left( \frac{1+x}{x} \right) ,
\eea
while $C_{s,a}^{i j}$ and $C_{s,b}^{i j}$ arise from self-energy corrections: 
\bea\label{csb}
C_{s,a}^{i j} (x)\ =\ \sqrt{x}\, C_{s,b}^{i j} (x)\ =\
  \frac{\sqrt{x}(1-x)}{(1-x)^{2}+ x \left(\lam \lam^{\dagger}\right)^2_{j j} / 16 \pi^2} \ .
\eea
The term proportional to $C_{s,b}^{i j}$ in Eq.~(\ref{eq:epsilon_I_deg})
does not contribute to the total CP asymmetry in $N_i$ decays
and is therefore a pure flavour
effect~\cite{CRV96}. In the limit of hierarchical right-handed neutrinos, $x\gg1$,
this term can be neglected and $C_{s,a}^{i j} (x) \simeq \sqrt{x} / (1-x)$. One then
recovers the non-resonant formula~(\ref{ECPI}).

The strength of the processes involving $\bar N_i$ ($\widetilde{N^c_i}$)
and $\ell_\alpha$ ($\tilde \ell_\alpha$) is parametrized by individual
washout parameters\footnote{We follow here the notations and conventions
of Refs.~\cite{matters,BdBP04}.} $\kappa_{i \alpha}$:
\beq
  \kappa_{i \al}\ \equiv\ \frac{\Gamma (N_i \rightarrow \ell_{\al} H_{u}) + \Gamma (N_i \rightarrow \bar \ell_{\al} H^*_{u})}{H(M_{i})}\ .
\eeq
Defining the effective neutrino masses
\bea
\tilde{m}_{i \al}\ \equiv\ \frac{|\lambda_{i \alpha}|^2 v_{u}^{2}}{M_{i}} \ ,
\eea
the individual washout parameters $\kappa_{i \al}$ can be expressed as
\bea
\kappa_{i \al}
= \frac{\tilde{m}_{i \al}}{m_{*}} \ ,
\eea
where $m_* = 16 \pi^{5/2}  \sqrt{g_{*}} v_{u}^{2} / (3 \sqrt{5} M_{P})
\simeq (1.56\times 10^{-3} \mbox{eV}) \sin^2 \beta$ is the equilibrium neutrino mass
(we used the fact that the number of effectively massless degrees of freedom
in the thermal bath at $T = M_1$ is $g_* = 228.75$ in the MSSM).
Summing over flavour indices, one obtains the total washout parameters $\kappa_i$: 
\begin{eqnarray}
\kappa_i\ =\ \sum_{\al} \kappa_{i \al}\ =\ \frac{\tilde{m}_i }{m_*}\ \simeq\
\frac{\tilde{m}_i }{(1.56\times10^{-3}\, {\rm eV}) \sin^2 \beta} \ .
\end{eqnarray}
The out-of-equilibrium condition for $N_i$ decays (resp. $N_i$ decays into
$\ell_\alpha$ or $\bar \ell_\alpha$) reads $\kappa_i < 1$ (resp. $\kappa_{i \alpha} < 1$).
If $N_i$ is in the strong washout regime ($\kappa_i \gg 1$), scattering processes
in the thermal bath produce an equilibrium population of $N_i$ at $T \sim M_i$.
When $T$ drops below $M_i$, the $N_i$ decay and generate asymmetries
in all lepton flavours, in proportions controlled by the $\epsilon_{i \alpha}$,
$\alpha= e, \mu, \tau$. Assuming that the decays occur in the temperature
regime where all flavours are relevant, the dynamical evolution of the individual
flavour asymmetries depends on the $\kappa_{j \alpha}$, where the index $j$
runs over the $N_j$ such that $M_j \lesssim M_i$. If $\kappa_{j \alpha} \gg 1$,
the asymmetry in the lepton flavour $\alpha$ is strongly washed out by the
lepton number violating processes involving $N_j$, which are in equilibrium
at $T = M_j$; if $\kappa_{j \alpha} \ll 1$, the opposite is true.
Consider now the case where $N_i$ is in the weak washout regime
($\kappa_i \ll 1$). Then its number density never reaches thermal equilibrium,
but since $\kappa_{i \alpha} \ll 1$ for all $\alpha$,
the individual flavour asymmetries are only weakly washed out
by $N_i$-related processes (they may however be
strongly washed out by $N_j$-related processes, $j \neq i$).

The evolution of the number densities is obtained by solving the set of Boltzmann
equations. As is usual done, we take into account the expansion of the universe
by defining comoving number densities $Y_{X} \equiv n_{X}/s$.
The supersymmetric Boltzmann equations for the heavy (s)neutrino comoving
number densities read~\cite{Plumacher97}:
\begin{eqnarray}
Y_{N_i}^{\prime} (z) & = & - 2\,\kappa_i \left( D_{i}(z)+S_{i}(z) \right)
  \left(Y_{N_i}(z)-Y_{N_i}^{eq}(z)\right) \ ,
\label{BEYN} \\
Y_{\tilde{N_i}}^{\prime}(z) & = & - 2\,\kappa_i \left( D_{i}(z)+S_{i}(z) \right)
  \left(Y_{\tilde{N_i}}(z)-Y_{\tilde{N_i}}^{eq}(z)\right) \ ,
\label{BEYNtilde}
\end{eqnarray}
where $Y_{\tilde{N_i}}$ stands for $Y_{\widetilde{N^c_i}} + Y_{\widetilde{N^c_i}^\dagger}$,
$z \equiv M_{1}/T$ and the symbol $^\prime$ stands for $d / dz$.
The equilibrium
densities appearing in Eqs.~(\ref{BEYN})
and ~(\ref{BEYNtilde}) are given by:
\bea
Y_{N_i}^{eq}(z) & = & \frac{135 \zeta(3)}{8 \pi^{4} g_{*}}\, R_{i}^{2}z^{2}K_{2}(R_{i} z)\
  \stackrel{T\gg M_{i}}{\longrightarrow}\ \frac{135 \zeta(3)}{4 \pi^{4} g_{*}}\
  \simeq\ 1.8\times 10^{-3} \ ,
\label{YNeq}  \\
Y_{\tilde N_i}^{eq}(z) & = & \frac{4}{3}\ Y^{eq}_{N_i}(z)\ ,
\label{YNtildeeq}
\eea
where $R_i \equiv M_i/M_1$, and we have corrected the high temperature behaviour
of the Maxwell-Boltzmann distribution by a factor of $3 \zeta (3) / 4$ in $Y_{N_i}^{eq}(z)$,
and by a factor of $\zeta (3)$ in $Y_{\tilde N_i}^{eq}(z)$~\footnote{Assuming
Maxwell-Boltzmann (MB) statistics as is customary, the equilibrium number density
at high temperature differs from the Fermi-Dirac (FD) and Bose-Einstein (BE) cases
by a numerical factor:
\beq
n_{_{FD}}(T\gg M)\ =\ \frac{3}{4} \, n_{_{BE}}(T\gg M)\
=\ \frac{3 \zeta(3)}{4}\,n_{_{MB}}(T\gg M) \ , \nonumber
\eeq
while $n_{_{FD}}(T) \simeq n_{_{BE}}(T) \simeq n_{_{MB}}(T)$ at low temperature
($T \ll M$).}.
The factor of $2$ in the right-hand side of Eqs.~(\ref{BEYN}) and~(\ref{BEYNtilde})
accounts for the fact that there are twice as many channels in the supersymmetric
case as in the non-supersymmetric case.
In the regime in which all lepton flavours are relevant, the individual
$\D_\al \equiv B/3-L_{\al}$ asymmetries are driven by
the following Boltzmann equations~\cite{matters,AKR06}:
\begin{eqnarray}
\label{beflav}
Y_{\D_\al}^{\prime}(z)&=&-2\sum_{i=1,2}\e_{i \al}\, \kappa_i \, \left( D_{i}(z)+S_{i}(z) \right) \left(Y_{N_i}(z)-Y_{N_i}^{eq}(z) + \left(Y_{ \tilde{N_i} }(z)-Y_{\tilde{N_i}}^{eq}(z)\right)\right) \nonumber \\ 
 &+& 2\sum_{i=1,2} \kappa_{i \al} \sum_{\be} W_{i}(z)\,A_{\al \be}\,Y_{\D_\be}(z)  \ ,
\end{eqnarray}
where $Y_{\D_\al}$ stands for the total $\Delta_\al$
asymmetry stored in the fermionic species and in their supersymmetric partners.
In Eqs.~(\ref{BEYN}), (\ref{BEYNtilde}) and~(\ref{beflav}), the thermally averaged
decay rates $D_i (z)$ are given by:
\beq
D_{i}(z)\ =\ R_{i}^{2} z\, \frac{K_{1}(R_{i} z)}{K_{2}(R_{i} z)} \ .
\eeq
The scatterings terms $S_i (z)$ account for Higgs-mediated $\D L=1$ scatterings
involving top quarks and antiquarks. They receive both s- and t-channel contributions: 
\beq
S_{i}(z)\ =\ 2 S_{s}^{i}(z) + 4 S_{t}^{i}(z) \ ,
\eeq
whose expression can be found in Ref.~\cite{BdBP04}.
The washout term $W_{i}(z) = W^{ID}_i (z) + W^S_i (z)$ results
from the contribution of inverse decays:
\beq
W_{i}^{ID}(z)\ =\ \frac{1}{4}R_{i}^{4} \, z^{3} K_{1}(R_{i} z) \ ,
\eeq
and $\D L=1$ scatterings~\cite{BdBP04}:
\beq
W_{i}^{S}(z)\ =\ \frac{W_{i}^{ID}(z)}{D_{i}(z)}\left(2 S_{s}^{i}(z)\left(\frac{Y_{N_i}(z)}{Y_{N_i}^{eq}(z)}+\frac{Y_{\tilde{N}_i}(z)}{Y_{\tilde{N}_i}^{eq}(z)}\right)+8 S_{t}^{i}(z)\right) .
\eeq

In writing the above Boltzmann equations, we made several assumptions
which we now proceed to clarify.
In the washout term, we neglected the off-shell part of the $\D L=2$ scatterings,
which is a good approximation as long
as $M_{i} \ll \kappa_{i \al}\, (10^{13} \GeV)$~\cite{matters}.
We also omitted $\D L=0$ scatterings such as $N_i N_j \rightarrow \ell \bar \ell$,
$N_i N_j \rightarrow H_u H^*_u$ and
$N_i \ell (\bar \ell) \rightarrow N_j \ell (\bar \ell)$, which do not contribute
to the washout but can affect the abundance
of the heavy (s)neutrinos (when flavour effects are taken into account, they
also tend to redistribute the lepton asymmetry among flavours).
These processes are of higher order in the neutrino Yukawa couplings
and are expected to have little impact on the final baryon asymmetry.
We further neglected the triplet-related washout processes,
gauge scatterings~\cite{GNRRS03,Pilaftsis}, spectator
processes~\cite{BP01}, and the higher order processes
{\small $1\rightarrow3$} and {\small $2\rightarrow3$}~\cite{NRR07}.
Finally, since the left-right symmetry is broken at a scale $v_R$ which may lie
several orders of magnitude below $M_{GUT}$, one may worry that decay and
scattering processes mediated by the heavy $SU(2)_R \times U(1)_{B-L}$
gauge bosons affect the heavy (s)neutrino number densities.
Indeed, $W_{R}\, $- and $Z^{\prime}$-mediated processes such as
$N_i\, e_R \rightarrow \bar q_R\, q'_R$ and $N_i N_i \rightarrow f \bar f$
tend to keep the heavy (s)neutrinos in thermal equilibrium, thus reducing
the generated lepton asymmetry~\cite{MSS98,CFL99}. As shown in
Ref.~\cite{Cosme04}, however, this effect can be practically neglected if
$M_i / v_R < 10^{-2}$, which turns out to be the case for $i=1,2$ in each
of the 8 solutions (at least as long as $N_{1,2}$ are light enough
to be thermally produced after reheating, i.e. $M_{1,2} \lesssim 10^{11}$ GeV).
Therefore, we do not need to include $W_{R}\, $- and $Z^{\prime}$-mediated
processes in our study\footnote{Note however that
$W_{R}\, $- and $Z^{\prime}$-mediated scatterings can help generating
an equilibrium population of $N_i$ in the weak washout regime ($\kappa_i \ll 1$),
provided that $M_i / v_R > 10^{-3}$~\cite{Cosme04}.}.

The Boltzmann equations~(\ref{beflav}) are written for the $B/3 - L_\alpha$
asymmetries $Y_{\D_\al}$ rather than for the lepton asymmetries $Y_{L_{\al}}$,
because the former are preserved by all MSSM interactions (including the
non-perturbative sphaleron processes), contrary to the latter.
As the washout term in Eq.~(\ref{beflav}) depends on $Y_{L_{\al}}$,
we need to express it in terms of the $Y_{\Delta_{\be}}$'s.
This is done by a conversion matrix $A$~\cite{BCST99}, whose entries
depend on which interactions are in equilibrium:
\beq
Y_{L_{\al}}\ =\ \sum_{\be} A_{\al \be}\, Y_{\D_\be}\ .
\eeq
Depending on the temperature at which leptogenesis takes place (identified
for simplicity with $M_1$ below), one must consider one of the following three regimes.
If $M_{1} \lesssim 10^{9} \GeV\, (1+\tan^2 \be)$, 
the tau and muon Yukawa interactions  are in equilibrium, hence all three
lepton flavours are distinguishable. One must then write three separate
Boltzmann equations for the flavour asymmetries $Y_{\D_e}$, $Y_{\D_\mu}$
and $Y_{\D_\tau}$, as in Eq.~(\ref{beflav}). The $3 \times 3$
$A$ matrix is given by~\cite{AKR06}:
\beq
A\ =\ \left( \begin{array}{ccc}
-93/110 & 6/55 &  6/55 \\ 
 3/40 &  -19/30 &  1/30 \\ 
 3/40 & 1/30 & -19/30
\end{array} \right) .
\label{bigA}
\eeq
If $10^{9} \GeV\, (1+\tan^2 \be) \lesssim M_1 \lesssim 10^{12} \GeV\, (1+\tan^2 \be)$, 
the muon Yukawa interactions are no longer in equilibrium, and the electron and muon
lepton flavours can no longer be distinguished. The lepton asymmetry must then be
projected onto the 2-flavour space $(Y_{L_{e+\mu}}, Y_{L_\tau})$,
where $Y_{L_{e+\mu}} \equiv Y_{L_e} + Y_{L_\mu}$, and correspondingly
$Y_{\D_{e+\mu}} \equiv Y_{\D_e} + Y_{\D_\mu}$. The conversion matrix
is now a $2 \times 2$ matrix:  
\beq
A\ =\ \left( \begin{array}{cc}
-541/761 & 152/761  \\ 
46/761 &  -494/761  
\end{array} \right) ,
\label{smallA} 
\eeq
and the Boltzmann equations for $Y_{\D_e}$ and $Y_{\D_\mu}$ must be replaced
by a single equation for $Y_{\D_{e+\mu}}$ involving the CP asymmetries
$\epsilon_{i, e+\mu} \equiv \epsilon_{ie}+\epsilon_{i\mu}$ and the washout
parameters $\kappa_{i, e+\mu} \equiv \kappa_{ie} + \kappa_{i\mu}$.
Finally, if $M_1 \gtrsim 10^{12} \GeV\, (1+\tan^2 \be)$, none of the interactions
involving charged lepton Yukawa couplings are in equilibrium, and one recovers
the flavour-independent treatment of leptogenesis with $A=-\unit$.

It has been pointed out in Ref.~\cite{BdBR06} that the out-of-equilibrium condition
used above to determine the flavour regime is actually not sufficient.
For a given flavour asymmetry $Y_{\D_\alpha}$ to evolve independently
during leptogenesis, the corresponding charged lepton Yukawa interaction rate
$\Gamma_{\alpha}(T) \simeq 5\times 10^{-3}\, h_{\alpha}^2 (1+\tan^2 \beta)\, T$
should not only be faster than the expansion rate of the Universe, but,
more importantly, it should also be faster than the
$N_i$ inverse decay rate. If this were not the case,
inverse decays would keep the evolution of the lepton state coherent.
For the tau and muon
lepton flavours, this condition reads
$M_1\, \lesssim\, 10^{12} \GeV\, (1 + \tan^2 \be) / \kappa_{1 \tau}$ and
$M_1\, \lesssim\, 10^9 \GeV\, (1 + \tan^2 \be) / \kappa_{1 \mu}$,
respectively. It is more restrictive than the out-of-equilibrium condition
for $\kappa_{1\tau} \gg 1$ (resp. $\kappa_{1\mu} \gg 1$).

The final baryon asymmetry is given by:
\beq
Y_{B}\ =\ \frac{10}{31} \sum_{\al} Y_{\D_\al}\ ,
\label{eq:conversion}
\eeq
where the factor $10/31$ is due to the partial conversion of the $\D_\al$
asymmetries into a baryon asymmetry by the non-perturbative
sphaleron processes~\cite{KS88} (we assume here that
sphalerons come out of thermal equilibrium below the electroweak
phase transition).  It has been pointed out recently~\cite{CGT08}
that the conversion factors relating the flavour asymmetries
$Y_{\Delta_\alpha}$ to $Y_B$ depend on the actual superpartner
spectrum. Strictly speaking, Eq.~(\ref{eq:conversion})
is only valid if all sfermions are heavy, while the final baryon asymmetry
can be reduced by a factor of $2/3$ if sleptons are light, as in some
minimal supergravity scenarios. Since we do not assume any particular
superpartner spectrum in this paper, we shall stick to Eq.~(\ref{eq:conversion})
in the following.

Being supersymmetric, the Grand Unified models we are considering
in this paper face the so-called gravitino
problem~\cite{gravitino_problem}: in an inflationary universe,
gravitinos are abundantly produced by thermal scatterings
in the reheating phase and, if they are unstable, their late decays
tend to spoil the successful predictions of Big Bang nucleosynthesis.
The requirement that this does not happen puts an upper bound
on the reheating temperature, $T_{RH} \lesssim (10^5 - 10^{10})$ GeV
for $300 \GeV \lesssim m_{3/2} \lesssim 30 \TeV$,
depending on the gravitino mass and on the superparticle
spectrum~\cite{KKMY08}. If instead the gravitino is the lightest
supersymmetric particle (LSP), the requirement
that its relic density does not exceed the dark matter abundance
leads to a weaker constraint, $T_{RH} \lesssim (10^9 - 10^{10})$ GeV
for a gravitino of mass $m_{3/2} \sim 100$ GeV~\cite{thermal_production}.
These constraints are in conflict with successful thermal leptogenesis,
which requires $M_1 \gtrsim 10^9$ GeV
in the type I seesaw case~\cite{DI02}.
In the left-right symmetric seesaw case studied in this paper,
we also find such a tension.
We nevertheless stick to thermal leptogenesis, since some
supersymmetric scenarios can accommodate a reheating temperature
$T_{RH} \sim 10^{10}$ GeV (see the discussion at the end of
Section~\ref{sec:dependence}).
In our numerical computation of the baryon asymmetry, we do not
explicitly include the dynamics of reheating in the Boltzmann equations,
as was done in Refs.~\cite{GNRRS03,AT06,HWP08},
but we take the reheating temperature into account in the initial conditions.
Namely, we start evolving the Boltzmann equations at some temperature
$T_{in}$, which we identify with $T_{RH}$. Heavy (s)neutrinos
with masses $M_{i} \gtrsim (4-5)\, T_{in}$ will thus give a negligible
contribution to the lepton asymmetry, because their production processes
(inverse decays and $\Delta L = 1$ scatterings) are Boltzmann-suppressed.

\section{Numerical results}
\label{sec:results}

In this section, we present our numerical results for leptogenesis
in the class of supersymmetric $SO(10)$ models described in
Section~\ref{sec:framework}. The final baryon asymmetry is
obtained by numerically integrating the Boltzmann
equations written in Section~\ref{sec:leptogenesis}, starting
the evolution at $T_{in} = 10^{11}$ GeV with vanishing abundances
for $N_{1,2}$ and $\tilde N_{1,2}$. With this choice of $T_{in}$
and $\tan \beta = 10$, one can consider that leptogenesis always
takes place in the 3-flavour regime. In Section~\ref{sec:dependence},
we shall investigate the effect of lowering $T_{in}$.

We present results for the following four reference solutions:
\begin{itemize}
\item $(+,+,+)$: this solution, which corresponds to dominance of the
type II seesaw contribution
in the large $v_R$ region, is characterized by a mild hierarchy
of right-handed neutrino masses, with $M_{1,2}$ growing with $v_R$.
\item $(+,-,+)$: this solution is characterized by an intermediate value
of $M_1$ ($M_1 \sim10^{9-10}$ GeV).
\item $(-,-,-)$: this solution, which reduces to the type I seesaw case
in the limit $v_R \rightarrow \infty$, is characterised by a hierarchical
right-handed neutrino mass spectrum with a small value of $M_1$
($M_1 \sim10^{4-5}$~GeV) and an intermediate value of $M_2$.
\item $(+,+,-)$: this solution differs from the previous one by the fact that
$M_2$ grows with $v_R$ rather than assuming a constant value.
\end{itemize}
The four remaining solutions show very similar patterns of right-handed
neutrino masses (the main differences are in the behaviour of $M_2$
and $M_3$ in the large $v_R$ region) and give analogous results
for the baryon asymmetry.
The input values for the quark masses and
mixing angles and for the measured neutrino oscillation parameters are
chosen as specified in Section~\ref{sec:framework}. We further assume
a hierarchical light neutrino mass spectrum with $m_1 = 10^{-3}$ eV
and $\theta_{13} = 0$; in Section~\ref{sec:dependence},
we shall study the effect of varying these input parameters. Rather
than attempting to perform a scan over the large number of available
CP-violating phases, we choose specific values for these phases
in order to illustrate the typical behaviour of the different solutions.
We first show the results obtained under the assumption that $M_d = M_e$
holds at the GUT scale; in a second stage, we include the necessary
corrections to this mass relation and investigate their effects on the
right-handed neutrino mass spectrum and on the final baryon asymmetry.
Finally, we set $\beta / \alpha = 0.1$ for practical reasons of numerical integration. 
As explained in Section~\ref{subsec:HLS}, changing the value of $\beta / \alpha$
only amounts to shift the curves along the $v_R$ axis.

For convenience, we shall plot the absolute value of the baryon
asymmetry $|Y_B|$
rather than $Y_B$ itself. Since the sign of $Y_B$
can be reverted by simply changing the sign of all CP-violating phases,
we do not loose any information by doing so, at least to the extent
that the effect of the CKM phase (whose sign is determined experimentally)
can be neglected. In practice this will be the case in all examples studied
in this paper, because the CKM phase always appears in combination
with small quark mixing angles, and large high-energy phases are assumed
to be present.

\subsection{Relevance of flavour effects}
\label{subsec:flavours}

\begin{figure}
\begin{center}
\includegraphics[width=8.cm]{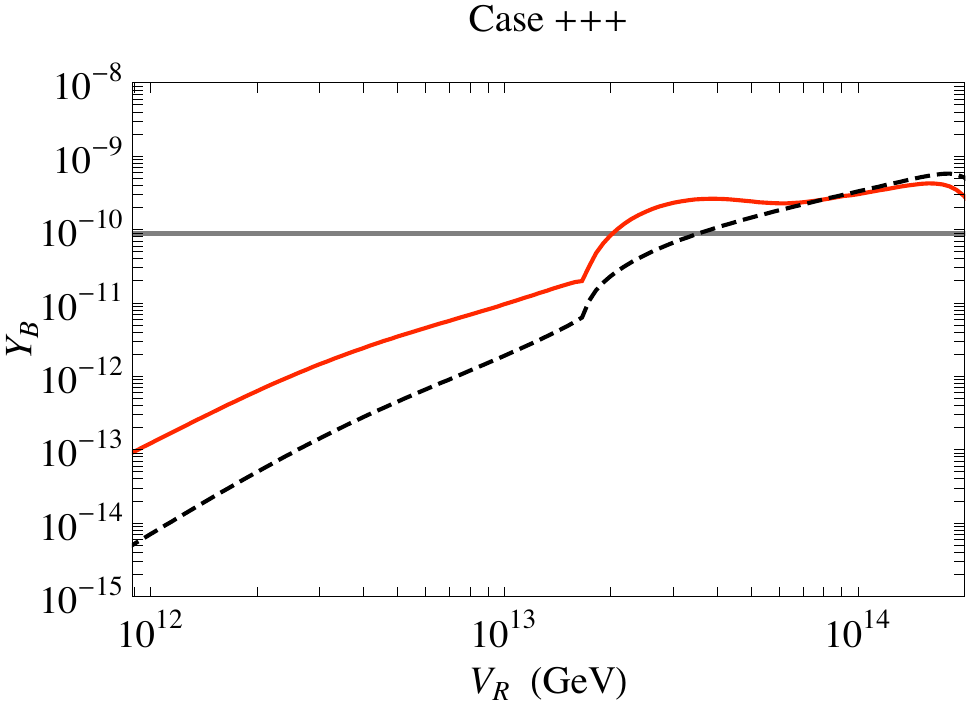} \hspace{0.5cm}
\includegraphics[width=8.2cm]{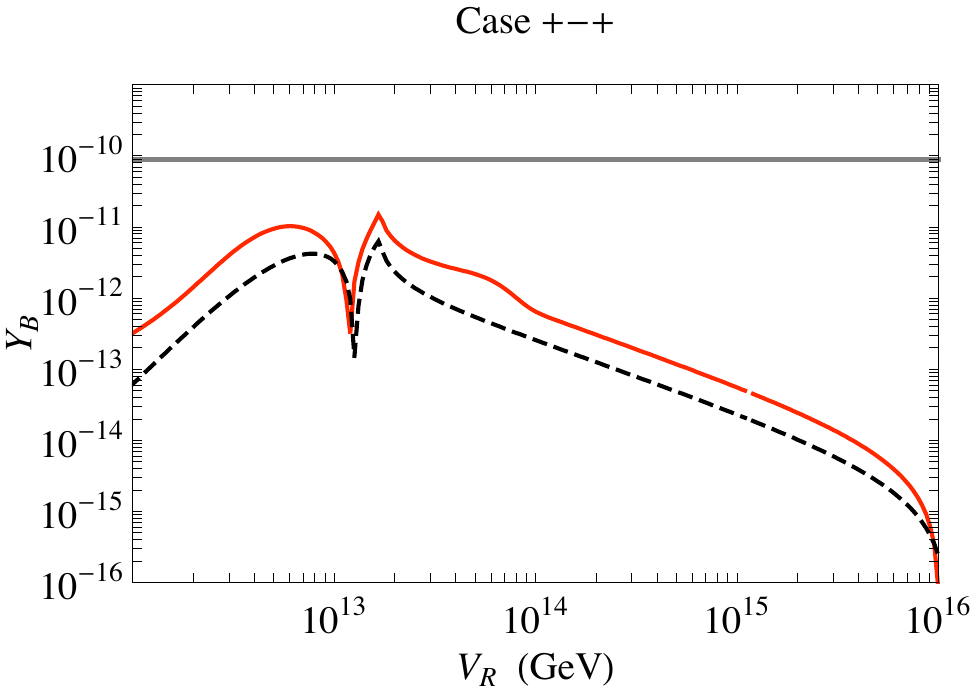}\\
\vspace{0.5cm}
\includegraphics[width=8.cm]{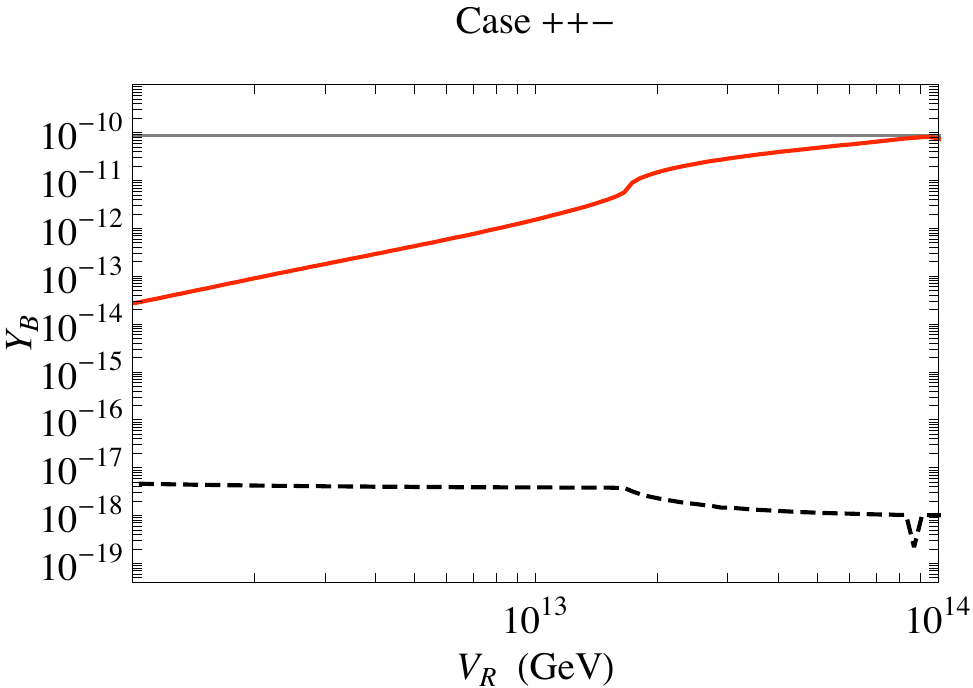} \hspace{0.5cm}
\includegraphics[width=8.cm]{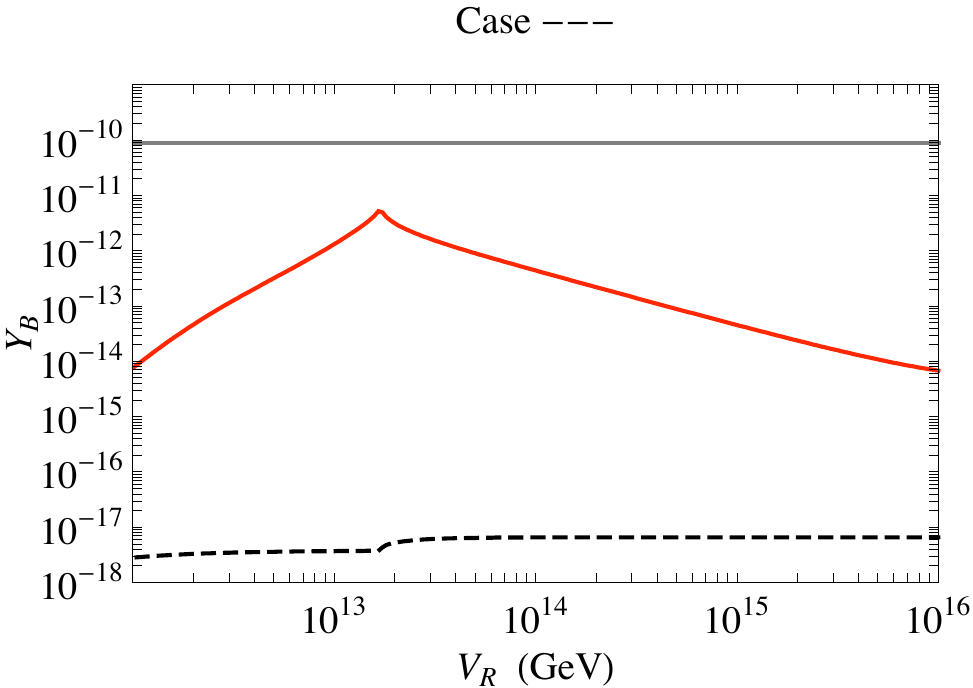}
\caption{The final baryon asymmetry as a function of $v_R$
for the four reference solutions, in the one-flavour approximation (dashed black line)
and with flavour effects taken into account (solid red line).
The GUT-scale mass relation $M_d = M_e$ is assumed.
Inputs: hierarchical light neutrino masses with $m_1 = 10^{-3}$ eV,
$\theta_{13} = 0$ and no CP violation in the PMNS mixing matrix;
$\Phi_2^u=\pi/4$ and all other high-energy phases are set to zero;
$\beta / \alpha = 0.1$. The Boltzmann equations are evolved starting
from $T_{in} = 10^{11}$ GeV. The thick horizontal line corresponds
to the WMAP constraint.}
\label{fig:Um=1}
\end{center}
\end{figure}

We first show in Fig.~\ref{fig:Um=1} the final baryon asymmetry in the absence
of corrections to the GUT-scale mass relation $M_d = M_e$, 
for the four reference solutions $(+,+,+)$, $(+,-,+)$, $(+,+,-)$ and $(-,-,-)$.
In order to estimate the relevance of flavour effects, we plotted
the result of the numerical computation both in the one-flavour approximation
(dashed black line) and in the flavour-dependent approach (solid red line).
One can see that flavour effects tend to enhance the baryon asymmetry
by up to one order of magnitude in the $(+,+,+)$ and $(+,-,+)$ cases.
Not surprisingly, solution $(+,+,+)$ leads to successful
leptogenesis for large values\footnote{We do not show the region
$v_R > 2 \times 10^{14}$ GeV, where $Y_B$ drops below the WMAP value.
This behaviour is due to the fact that, above
$v_R = \mbox{few} \times 10^{14}$ GeV, $N_1$ and $N_2$ are too heavy
to be thermally produced ($M_{1,2} \gg T_{in}$).} of $v_R$,
where $M_1 \gtrsim 10^{10}$ GeV; flavour effects allow this solution
to be successful for smaller values of $v_R$ (i.e. for smaller values of $M_1$)
than in the one-flavour approximation.
By contrast, solution $(+,-,+)$ fails to generate the observed baryon
asymmetry due to the strong washout by inverse decays and $\Delta L = 1$
scatterings, and this conclusion still holds for different choices
of the CP-violating phases.

Flavour effects have a much more dramatic impact in the
$(+,+,-)$ and $(-,-,-)$ cases, which are characterized by a strong hierarchy
between $M_1$ and $M_2$. In these solutions, the observed enhancement
of $Y_B$ is due to the fact that the asymmetry in a particular lepton flavour
is only mildly washed out by $N_1$-related processes, while the total washout
is strong. As a consequence, the asymmetry generated in $N_2$ decays
is completely washed out in the one-flavour approximation, while its projection
on this particular flavour survives when flavour effects are taken into
account. This effect, which has been first identified in the type I seesaw
framework in Ref.~\cite{Vives05},
will be discussed in greater detail in Subsection~\ref{subsec:N2}.
Despite the huge increase in $Y_B$, however, solution $(-,-,-)$
fails to reach the WMAP level, while solution $(+,+,-)$ is marginally
successful for $v_R \approx 10^{14}$ GeV, where $M_2 \sim T_{in}$
(for larger values of $v_R$, $M_2 \gg T_{in}$ and $N_2$ no longer
contributes to $Y_B$, which then drops well below the WMAP value).

\subsection{Relevance of the corrections to the mass relation $M_d = M_e$}
\label{subsec:corrections}

The results shown in Fig.~\ref{fig:Um=1}
were obtained assuming $M_d = M_e$ to hold at the GUT scale,
which only agrees at the order of magnitude level with the measured
down quark and charged lepton masses.
Let us now include the necessary corrections
to this relation and investigate their effects on the final baryon asymmetry.
Given the assumptions made about the Higgs sector in
Section~\ref{subsec:HLS}, these corrections must come from
non-renormalizable operators\footnote{One could alternatively relax
the assumptions made in Section~\ref{subsec:HLS} and introduce
a $\bf 210$ Higgs representation in order to generate vev's for
the $SU(2)_L$ doublet components of the $\bf \overline{126}$.
In this case, the $f_{ij}$ couplings would contribute both to
the left-right symmetric seesaw formula and to the down quark
and charged lepton masses, which would render their reconstruction
much more difficult.}. The simplest possibility is to add
the following terms to the superpotential (see e.g. Ref.~\cite{ARDHS93}):
\beq
  \frac{\kappa_{ij}}{\Lambda}\  {\bf 16}_i {\bf 16}_j {\bf 10}_1 {\bf 45}\ ,
\label{eq:nr_operators}
\eeq
where we assume that the $Y\! =\! +1$  Higgs doublet in ${\bf 10}_1$
does not acquire a vev, so that 
the mass relation $M_D = M_u$ is left untouched\footnote{If the operators
${\bf 16}_i {\bf 16}_j {\bf 10}_2 {\bf 45}$ were also present, they would give an
antisymmetric contribution to $M_D$ and the reconstruction procedure
would no longer be applicable. We assume here that
they are forbidden by some symmetry.}.
The operators (\ref{eq:nr_operators}) will modify
the reconstructed couplings $f_{ij}$ (hence the right-handed neutrino
masses and couplings) by introducing a mismatch $U_m$
between the bases of left-handed charged lepton and down quark mass eigenstates:
\beq
  M^\dagger_e M_e\ =\ \hat M^2_e\ , \qquad
  M^\dagger_d M_d\ =\ U^\dagger_m \hat M^2_d U_m\ , \qquad
  M^\dagger_u M_u\ =\ U^\dagger_m U^\dagger_q \hat M^2_u U_q U_m\ ,
\eeq
where the mass matrices are written
in the basis of charged lepton mass eigenstates, and $\hat M_e$, $\hat M_d$,
$\hat M_u$ are diagonal eigenvalue matrices.
Note that $M_d$ and $M_e$ are no longer symmetric in the original
basis, since they receive antisymmetric contributions from the operators
in Eq.~(\ref{eq:nr_operators}).
The measured down quark and charged lepton masses do not determine
completely the unitary matrix $U_m$, but constrain its mixing angles
to lie in restricted ranges (see the Appendix for details).
One of them can be taken to be either large
($\theta_{12}^m\sim1$) or somewhat smaller ($\theta_{12}^m\sim0.2-0.3$).

\begin{figure}[p]
\begin{center}
\includegraphics[width=7.7cm]{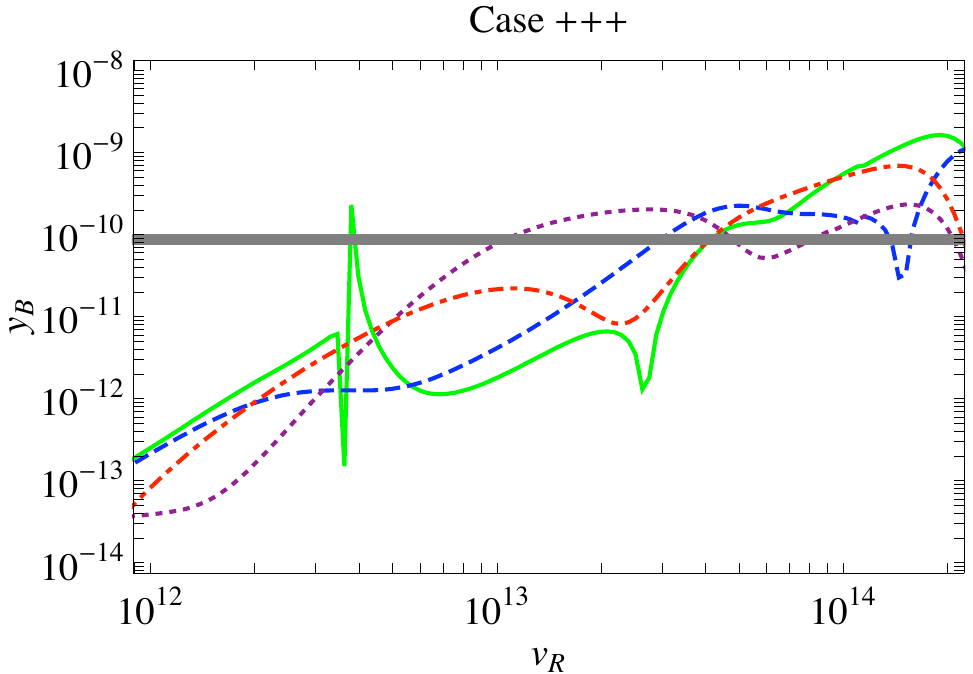}\hspace{0.7cm}
\includegraphics[width=7.5cm]{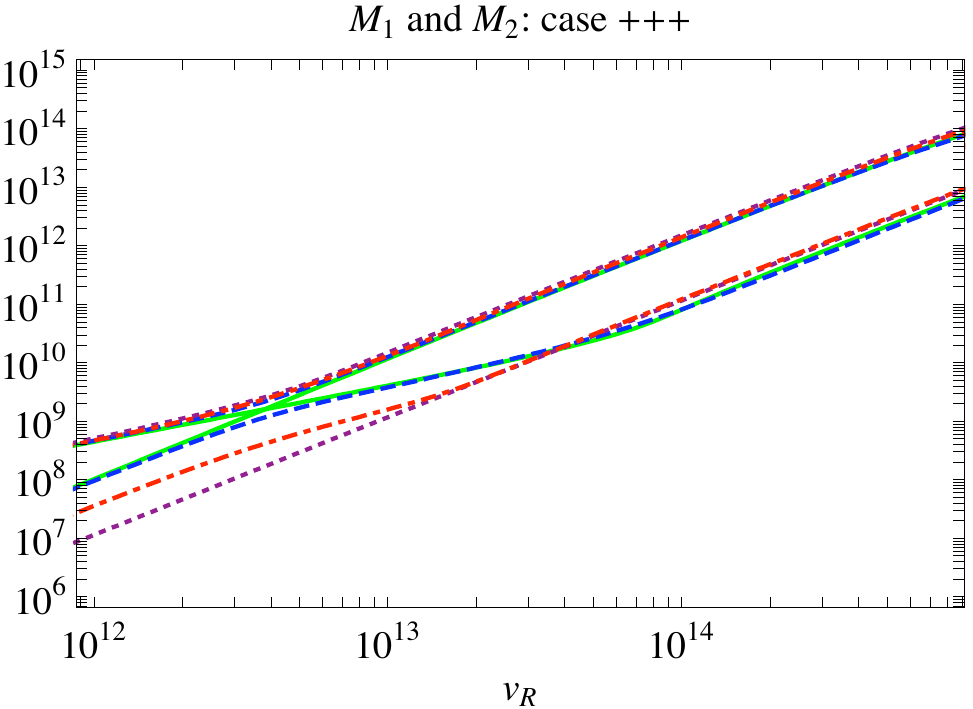}\\
\vspace{0.3cm}
\includegraphics[width=7.7cm]{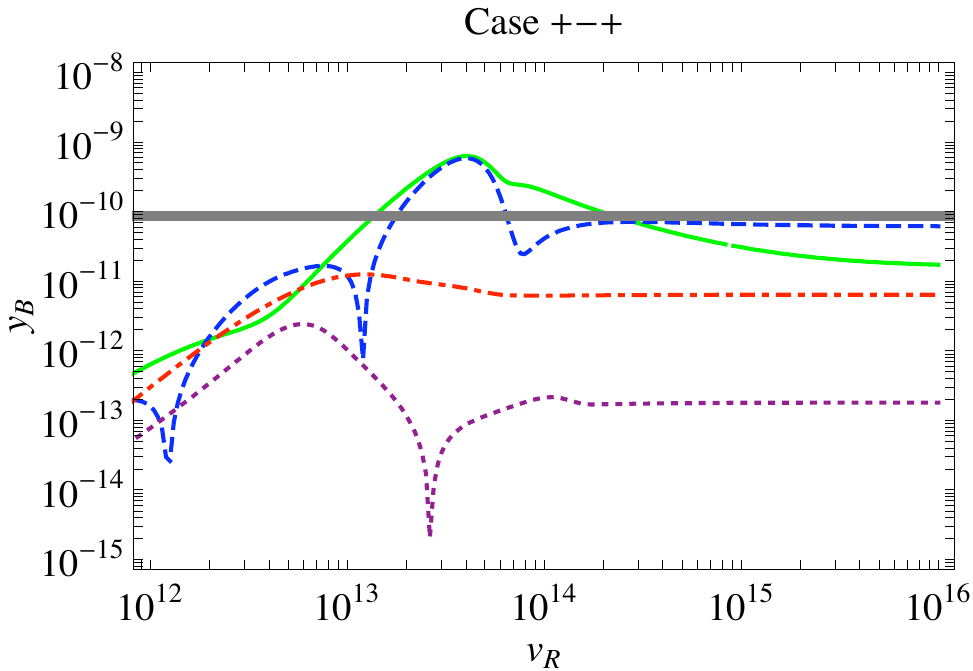}\hspace{0.5cm}
\includegraphics[width=7.5cm]{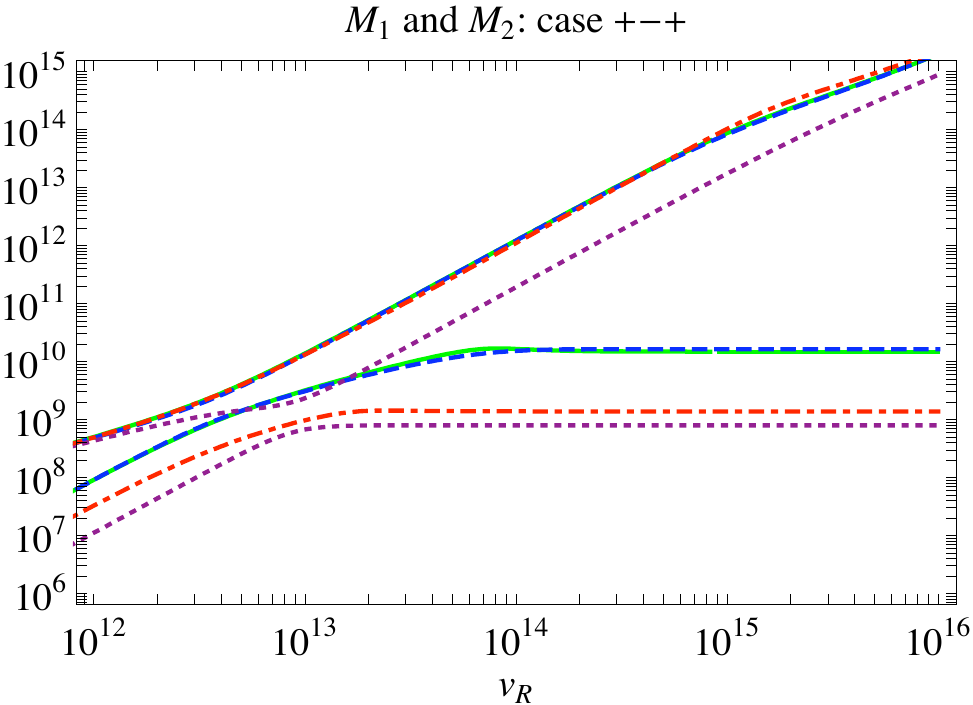}\\
\vspace{0.3cm}
\includegraphics[width=7.7cm]{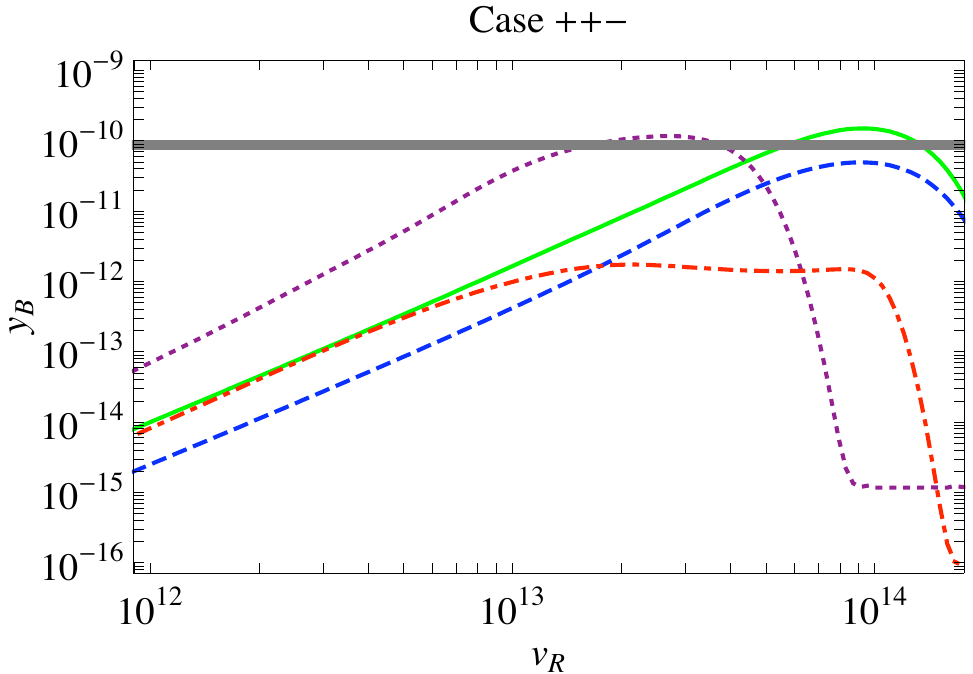}\hspace{0.7cm}
\includegraphics[width=7.5cm]{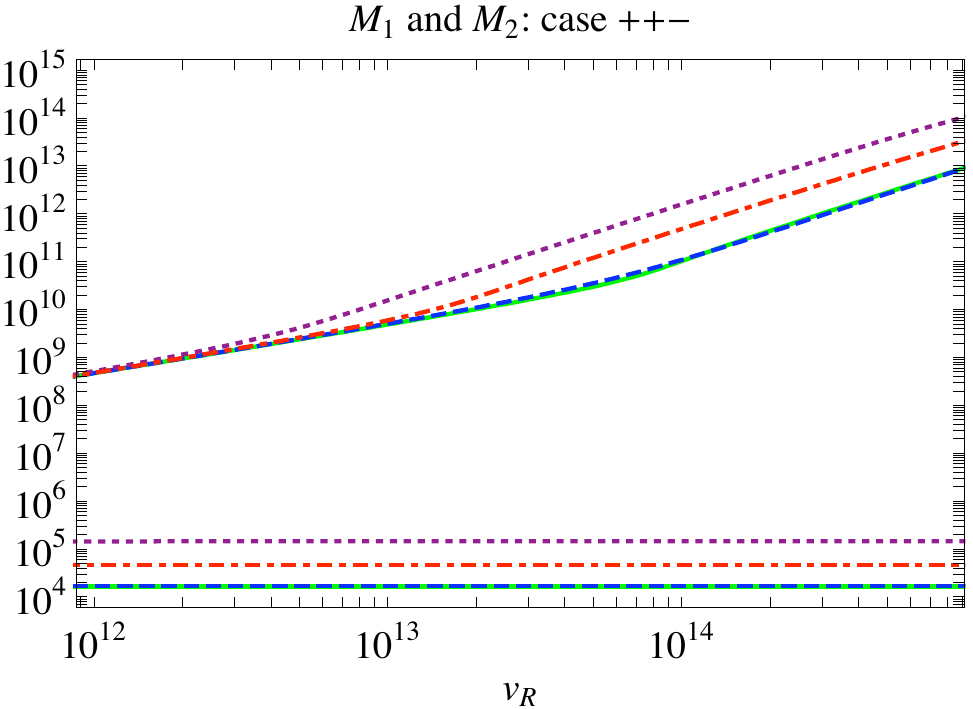}\\
\vspace{0.3cm}
\includegraphics[width=7.7cm]{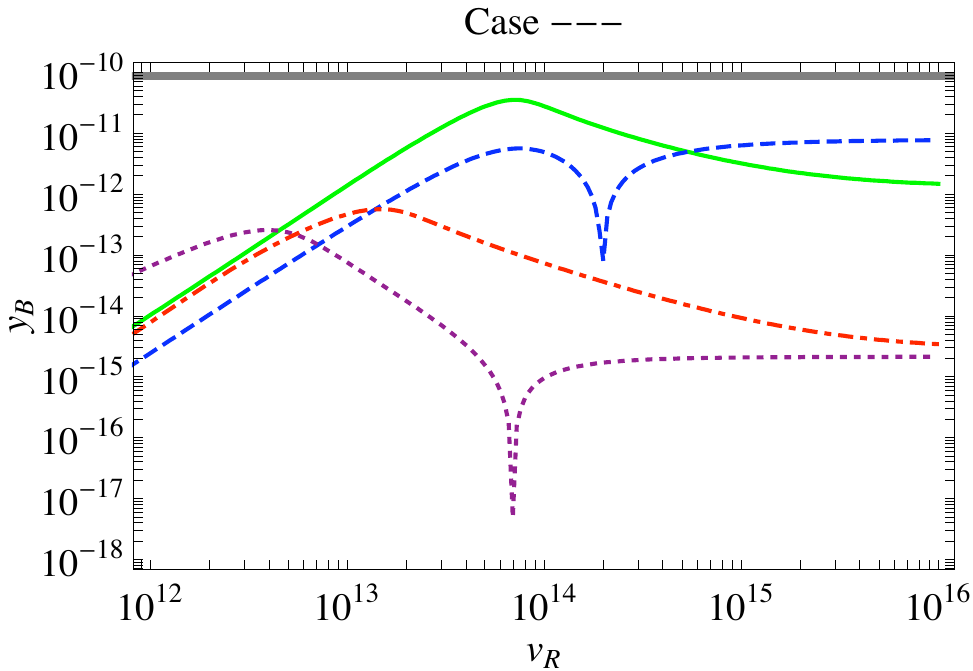}\hspace{0.5cm}
\includegraphics[width=7.5cm]{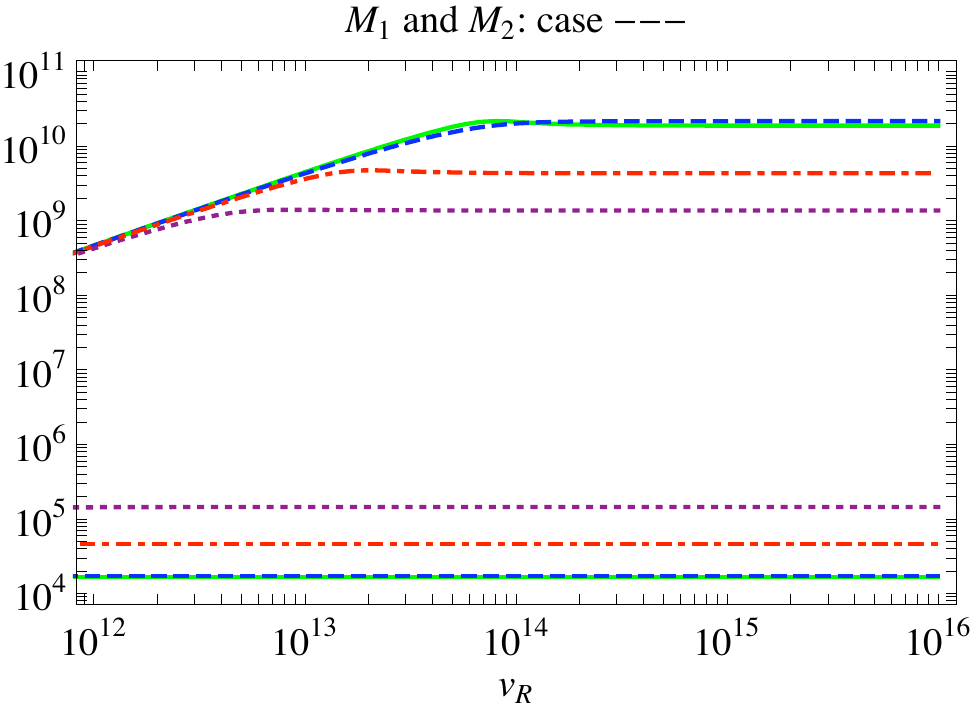}
\caption{Final baryon asymmetry (left panels) and masses of $N_1$ and $N_2$
(right panels) as a function of $v_R$ in the four reference solutions with a non-trivial
$U_m$ and a non-vanishing Majorana or high-energy phase.
The solid green, dashed blue, dotted purple and dash-dotted red lines
corresponds to the sets 1, 2, 3 and 4 described in the Appendix, respectively.
The other input parameters are as in Fig.~\ref{fig:Um=1}.}
\label{fig:Um}
\end{center}
\end{figure}

Fig.~\ref{fig:Um} shows the final baryon asymmetry for four representative
choices of $U_m$.
The main difference with the $M_d = M_e$ case is that several
choices for $U_m$ lead to successful leptogenesis in the solution $(+,-,+)$.
This is an interesting result, since this solution is special to the left-right
symmetric seesaw mechanism: it does not correspond to dominance
of either the type I  or the type II seesaw mechanism in the light neutrino
mass matrix in the large $v_R$ limit. As for solution $(-,-,-)$,
we were unable to find values of $U_m$ and of the Majorana and
high-energy phases allowing the final baryon asymmetry to reach
the observed value. There is a general tendency for $Y_B$ to reach
larger values for intermediate values of $v_R$
(typically $10^{13} \GeV \lesssim v_R \lesssim 10^{14} \GeV$ for our
choice $\beta / \alpha = 0.1$), where type I and type II contributions
partially compensate for each other in the light neutrino mass
matrix.

The enhancement of $Y_B$
observed for some choices of $U_m$ can be explained by the influence
of $\theta_{12}^m$ on the CP asymmetries $\epsilon_{i \alpha}$ and on the
washout parameters $\kappa_{i\alpha}$ (see the discussion at the end of
the Appendix).
In particular, solution $(+,-,+)$ is found to be successful for large
values of $\theta_{12}^m$, as in the sets 1 and 2 of the Appendix.
We note in passing that a large mixing in $U_m$ also
enhances lepton flavour violation, so that processes like $\mu\to e\gamma$
or $\tau\to\mu\gamma$ might be close to their experimental limits (and
a significant portion of the supersymmetric parameter space is already
excluded).
To illustrate the effect of $U_m$ on the right-handed neutrino mass spectrum, 
we also plotted $M_1$ and $M_2$ as a function of $v_R$  in Fig.~\ref{fig:Um}.
The enhancement of $Y_B$ with respect to the $M_d = M_e$ case
in solutions $(+,-,+)$ and $(-,-,-)$ is correlated with, respectively, an increase
of $M_1$ by a factor of 10, and an increase of $M_2$ by a factor of 5.
One can also notice that successful leptogenesis is generally associated
with right-handed neutrino masses above $10^{10}$ GeV, which indicates
a conflict with the gravitino problem. Some of the solutions that are successful
for $T_{in} = 10^{11}$ GeV might actually fail for $T_{in} < 10^{10}$ GeV,
and we can already anticipate that this will be the case for solution $(+,+,-)$.
We shall come back to this point in Section~\ref{sec:dependence},
where the dependence of $Y_B$ on
the reheating temperature will be discussed.

\subsection{Flavour effects and $N_2$ leptogenesis}
\label{subsec:N2}

Let us now discuss in more quantitative terms 
the interplay of flavour effects and $N_2$ decays
in the four solutions $(\pm,\pm',-)$, which are characterized by a strong hierarchy
between the masses of the two lightest right-handed neutrinos.
As mentioned above,
the lepton asymmetry generated by $N_2$ is exponentially
washed out by $N_1$-related processes in the one-flavour approximation.
Since $N_1$ has a small coupling to a particular lepton flavour,
however, the asymmetry in this flavour is only mildly washed out
in the flavour-dependent  treatment, and this explains the spectacular
enhancement of $Y_B$ observed in Fig.~\ref{fig:Um=1}.
We shall refer to this situation as ``flavour-dependent
$N_2$ leptogenesis'' in the following.

For flavour-dependent $N_2$ leptogenesis to be possible,
some conditions on the CP asymmetries and washout parameters
must be satisfied.  In Table~\ref{tab:eps-kappa}, we list the values
of $\epsilon_{i\al}$ and $\kappa_{i \al}$ ($i=1,2$; $\al = e, \mu, \tau$)
in the $(-,-,-)$ and $(+,+,-)$ solutions for $v_R = 10^{14}$ GeV, assuming
the same input parameters as in Fig.~\ref{fig:Um=1} (in particular,
$U_m = \unit$, $\Phi_{2}^{u}=\pi/4$ and all other CP-violating phases
are set to zero).
\begin{table}
\begin{center}
\begin{tabular}{|c|c|c|c|c|c|c|}
\hline $(-,-,-)$ & $N_{1}, e$ & $N_{1}, \mu$ & $N_{1}, \tau$
& $N_{2}, e$ & $N_{2}, \mu$ & $N_{2}, \tau$ \\ 
\hline $\e_{i\al}$ & $1.1 \times 10^{-16}$ & $9.6 \times 10^{-15}$
& $5.8 \times 10^{-14}$ & -$1.2 \times 10^{-7}$ & -$6.4 \times 10^{-8}$
& -$3.4 \times 10^{-7}$ \\ 
\hline $\kappa_{i\al}$ & 0.04 & 17.2 & 16.2 & 2.3 & 0.7 & 2.7 \\ 
\hline $(+,+,-)$ & $N_{1}, e$ & $N_{1}, \mu$ & $N_{1}, \tau$
& $N_{2}, e$ & $N_{2}, \mu$ & $N_{2}, \tau$ \\  
\hline $\e_{i\al}$ & $1.2 \times 10^{-16}$ & $9.7 \times 10^{-15}$
& $5.7 \times 10^{-14}$ & $7.0 \times 10^{-7 \,} $& $2.0 \times 10^{-7 \,}$
& $2.6 \times 10^{-6 \,}$ \\ 
\hline $\kappa_{i\al}$ & 0.04 & 17.2 & 16.2 & 0.5 & 0.2 & 3.5 \\ 
\hline 
\end{tabular}
\end{center}
\caption{Values of $\epsilon_{i\al}$ and $\kappa_{i \al}$ ($i=1,2$; $\al = e, \mu, \tau$)
in solutions $(-,-,-)$ and $(+,+,-)$. The input parameters are chosen as in
Fig.~\ref{fig:Um=1}, and the $B-L$ breaking scale is $v_R =10^{14}$ GeV.}
\label{tab:eps-kappa}
\end{table}
Both solutions exhibit a similar pattern of flavoured parameters:
the CP asymmetries in $N_1$ decays are extremelly small
($\epsilon_{1\alpha} < 10^{-13}$), and the washout induced by $N_1$
is strong except for the electron flavour ($\kappa_{1e} = 0.04$,
while $\kappa_{1\mu} \approx \kappa_{1\tau} \approx 16$).
By contrast, the CP asymmetries in $N_2$ decays are in the
$(10^{-7} - 10^{-6})$ range, and the $N_2$-induced washout
is moderate.

\begin{figure}
\begin{center}
\includegraphics[scale=0.5]{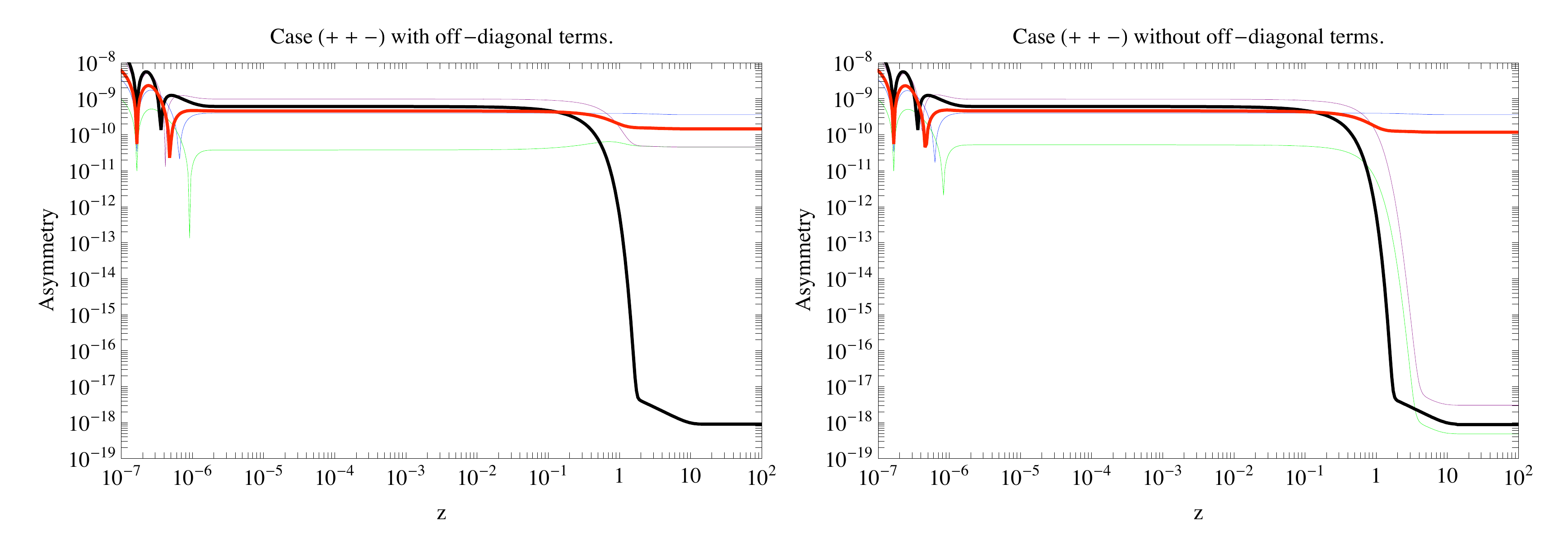}
\caption{Evolution of the asymmetries as a function of $z=M_1/T$ in the $(+,+,-)$
solution, with the off-diagonal entries of the $A$ matrix included (left panel)
and omitted (right panel).
The thin lines represent the three lepton flavour asymmetries: $Y_{\Delta_e}$ in blue
(medium grey), $Y_{\Delta_\mu}$ in green (light grey), $Y_{\Delta_\tau}$ in purple
(dark grey), while the thick red (medium grey) and black lines
stand for $Y_{B}$ and $Y^{\scriptscriptstyle 1FA}_{\scriptscriptstyle B-L}$,
respectively. The input parameters are chosen as in Fig.~\ref{fig:Um=1},
and the $B-L$ breaking scale is $v_R =10^{14}$ GeV.}
\label{fig:N2leptogenesis}
\end{center}
\end{figure}

Let us see in detail how these features explain the results observed
in Fig.~\ref{fig:Um=1}, focusing on solution $(+,+,-)$ for definiteness
(for earlier discussions of flavour-dependent $N_2$ leptogenesis
in the type I seesaw framework, see Refs.~\cite{Vives05} and~\cite{SY07}).
$N_2$ decays first generate asymmetries $(Y_{\Delta_\alpha})_{_{N_2}}$
in all three lepton flavours. Since $M_1 \ll M_2$, the processes involving $N_1$
are out of equilibrium at $T \sim M_2$, and the $(Y_{\Delta_\alpha})_{_{N_2}}$
can be computed taking into account $N_2$-induced washout only.
Neglecting off-diagonal entries in the $A$ matrix, one obtains:
\beq
  (Y_{\Delta_e})_{_{N_2}} \simeq\, -4\times 10^{-10}\ , \quad
  (Y_{\Delta_\mu})_{_{N_2}}\simeq\, -4\times 10^{-11} \ , \quad
  (Y_{\Delta_\tau})_{_{N_2}} \simeq\, - 10^{-9}\ ,
\eeq
while in the one-flavour approximation the $B-L$ asymmetry
induced by $N_2$ is $(Y^{\scriptscriptstyle 1FA}_{\scriptscriptstyle B-L})_{_{N_2}} \simeq - 6\times 10^{-10}$.
As the Universe cools down,
$N_{1}$-related washout processes come into equilibrium,
and the evolution of the $Y_{\Delta_\alpha}$'s is then governed
by the following Boltzmann equations:
\beq
Y_{\Delta_\al}^{\prime}(z)\ =\ 2 \kappa_{1\al} A_{\al \al}W_{1}(z)Y_{\Delta_\al}(z)
  + 2 \kappa_{1\al} \sum_{\be\neq\al}A_{\al \be}W_{1}(z)Y_{\Delta_\be}(z) \ ,
\label{eq:BE_N2lepto}
\eeq
in which the source term proportional to $\epsilon_{1\alpha}$ has been
neglected because of its smallness.
Eq.~(\ref{eq:BE_N2lepto}) yields the  formal solution:
\beq
Y_{\Delta_\alpha}(z)\ \simeq\ (Y_{\Delta_\alpha})_{_{N_2}}\, e^{2 A_{\al \al}
  \kappa_{1\alpha}\int_{zin}^{z}dx\  W_{1}(x)}\,
 +\, 2 \kappa_{1\alpha}\sum_{\be \neq \al}A_{\al \be} \int_{zin}^{z} dx\,
 W_{1}(x)Y_{\Delta_\be}(x)\,e^{2 A_{\al \al} \kappa_{1\alpha}\int_{x}^{z}dy \ W_{1}(y)} \ ,
\eeq
where the first term corresponds to the depletion of $Y_{\Delta_\al}$
due to $N_1$-related washout processes, whereas the second term
represents the effect of the flavour mixing induced by the off-diagonal
entries in the $A$ matrix.
Neglecting the off-diagonal entries of the $A$ matrix for the moment,
and omitting for simplicity the scattering terms in $W_1(z)$, one obtains:
\beq
Y_{\Delta_\alpha}^{d}\ \simeq\ e^{\frac{3 \pi}{4} A_{\al \al} \kappa_{1\alpha}}\,
  (Y_{\Delta_\alpha})_{_{N_2}}\ ,
\eeq
where we have used $\int_0^\infty dz\, z^3 K_1(z) = 3 \pi / 2$.
Since $\kappa_{1e} \ll 1 \ll \kappa_{1\mu (\tau)}$, the asymmetry
in the electron flavour is almost unaffected by $N_1$-induced washout,
while $(Y_{\Delta_\mu})_{_{N_2}}$ and $(Y_{\Delta_\tau})_{_{N_2}}$
are exponentially diluted, namely by a factor of order $10^{-11}$.
The final baryon asymmetry is\footnote{As explained earlier in this section,
the sign of $Y_B$ is not relevant since it can be reverted by changing
the sign of $\Phi^u_2$ (if one neglects the small contribution
of $\delta_{CKM}$ to $Y_B$).}:
\beq
Y_{B}\ \simeq\ \frac{10}{31}\, Y^d_{\Delta_e}\
  \simeq\  \frac{10}{31}\, 0.92\, (Y_{\Delta_e})_{_{N_2}}\ \simeq\ -1.2 \times 10^{-10} \ ,
\eeq
in good agreement with the numerical result. In the one-flavour approximation instead,
the $B-L$ asymmetry generated in $N_2$ decays is completely washed out by
$N_1$-related processes:
\beq
Y^{\scriptscriptstyle 1FA}_{\scriptscriptstyle B-L}\ \simeq\ e^{- \frac{3 \pi}{4} \kappa_1}\,
  (Y^{\scriptscriptstyle 1FA}_{\scriptscriptstyle B-L})_{_{N_2}}\
  \simeq\ 6 \times 10^{-35}\, (Y^{\scriptscriptstyle 1FA}_{\scriptscriptstyle B-L})_{_{N_2}}\ ,
\eeq
so that the dominant contribution to
$Y^{\scriptscriptstyle 1FA}_{\scriptscriptstyle B-L}$ actually comes from $N_1$
decays, in spite of the smallness of $\epsilon_1$ (an analogous statement can
be made about $Y_{\Delta_\mu}$ and $Y_{\Delta_\tau}$ in the flavour-dependent
treatment). All these results are illustrated in the right panel of
Fig.~\ref{fig:N2leptogenesis}.

Let us now add the effect of the off-diagonal entries in the $A$ matrix.
The contribution to $Y_{\Delta_\alpha}$ of the second term in the right-hand
side of Eq.~(\ref{eq:BE_N2lepto}) has been evaluated in Ref.~\cite{FXASMAA},
in the non-supersymmetric case:
\beq
Y_{\Delta_\alpha}^{od}\ \simeq\ 
  \frac{1.3\, \kappa_{1\al}}{1+0.8 (-A_{\al \al} \kappa_{1\al})^{1.17}}\,
\sum_{\be\neq \al} A_{\al \be} Y_{\Delta_\be}^{d}\ .
\eeq
This flavour mixing does not affect $Y_{\Delta_e}$, but
prevents the complete depletion of $Y_{\Delta_\mu}$ and $Y_{\Delta_\tau}$:
\bea
Y_{\Delta_{\mu (\tau)}}^{od}\ \simeq\ 0.12\,Y^d_{\Delta_e}\ \simeq\ -4.4\times 10^{-11}\ .
\eea
The final baryon asymmetry is only marginally affected,
reaching $Y_B \simeq - 1.5 \times 10^{-10}$. These analytic estimates are
confirmed by the numerical results shown in the left panel of
Fig.~\ref{fig:N2leptogenesis}.

We conclude that flavour effects play a crucial role in the four solutions $(\pm, \pm',-)$,
where they render $N_2$ leptogenesis possible. In spite of the spectacular
enhancement of the final baryon asymmetry with respect to the one-flavour
approximation, however, successful flavour-dependent $N_2$ leptogenesis
is difficult to achieve, as shown by Figs.~\ref{fig:Um=1} and~\ref{fig:Um}.
This is even more so in solution $(-,-,-)$, which in the large $v_R$ limit reduces
to the type I seesaw case, in which flavour-dependent $N_2$ leptogenesis
was originally proposed as a way to achieve
successful baryogenesis in GUTs~\cite{Vives05}.

\section{Dependence on the input parameters}
\label{sec:dependence}

The results presented in the previous section were obtained for
fixed values of the light neutrino parameters: we assumed a normal
mass hierarchy with $m_1 = 10^{-3}$ eV, $\theta_{13} = 0$ and no CP
violation in the PMNS mixing matrix, while
the oscillation parameters were taken from Ref.~\cite{FLMP05}.
Furthermore, the relation $M_D = M_u$ was assumed and the Boltzmann
equations were evolved from $T_{in} = 10^{11}$ GeV. In this section,
we numerically study the influence of these input parameters on the final baryon
asymmetry.

\subsection{Dependence on the light neutrino parameters}

Let us first study
the impact of the yet unmeasured light neutrino parameters:
the lightest neutrino mass, the mixing angle $\theta_{13}$ and
the Dirac phase $\delta_{PMNS}$, and finally
the type of mass hierarchy.

\subsubsection{Lightest neutrino mass (normal mass hierarchy)}

Since the flavour-dependent CP asymmetries $\epsilon_{1 \alpha}$ are bounded by
(note that the upper bound is the same as in the type I case~\cite{matters}):
\beq
  |\epsilon_{1 \alpha}|\ \leq\ \epsilon^{max}_{1\alpha}\ \equiv\
  \frac{3}{8 \pi}\, \frac{M_1 m_{max}}{v^2_u}\, \sqrt{\frac{\kappa_{1\alpha}}{\kappa_1}}\ , 
\eeq
and the type I inequality $\kappa_1 \geq m_1 / m_\star$ does not hold, one may
expect that successful leptogenesis is easier to achieve for quasi-degenerate
light neutrinos,  $m_1 \gtrsim 0.1$ eV. However, varying $m_1$ also modifies
the reconstructed right-handed neutrino parameters, which in turn affects the
CP asymmetries and washout parameters. In particular, the right-handed
neutrino masses are modified by an increase of $m_1$ in the following way
(the right-handed neutrino couplings $\lambda_{i\alpha}$ are also affected
via the $U_f$ matrix): 
the $M_i$'s belonging to
a ``type I branch'' decrease, while the ones belonging to a ``type II branch'' rise.
This behaviour can be understood by noticing that, to a good approximation,
the right-handed neutrino masses are proportional to the $x_i$'s
(see the Appendix B of Ref.~\cite{HLS06}). Since raising $m_1$ increases
the value of the $z_i$'s, Eq.~(\ref {eq:x_pm_limit}) implies that
the $M_i$ associated with some $x^-_j$ (``type I branch'') decreases,
while the $M_i$ associated with some $x^+_j$ (``type II branch'')
shows the opposite behaviour\footnote{Strictly speaking, this is only true
for $4 \alpha \beta \ll |z_j|^2$ ($v_R \gg 2 \sigma_u v^4_u / M_\Delta |z_j|^2$).
In the opposite limit (which is not relevant for the discussion below),
$x_j$ is almost independent of $z_j$ and the associated $M_i$ is
not affected by an increase of $m_1$.}.
These considerations explain to a large extent the impact of a variation
of $m_1$ on the final baryon asymmetry, since the upper bound on
$\epsilon_{1 \alpha}$ is proportional to $M_1$ and
$\kappa_{1\alpha} \propto 1 / M_1$: an increase in $M_1$ tends to
enhance the CP asymmetry in $N_1$ decays and to reduce the
$N_1$ washout parameters. An analogous statement can be made
about $M_2$ and the $N_2$-related leptogenesis parameters.

\begin{figure}
\begin{center}
\includegraphics[scale=0.6]{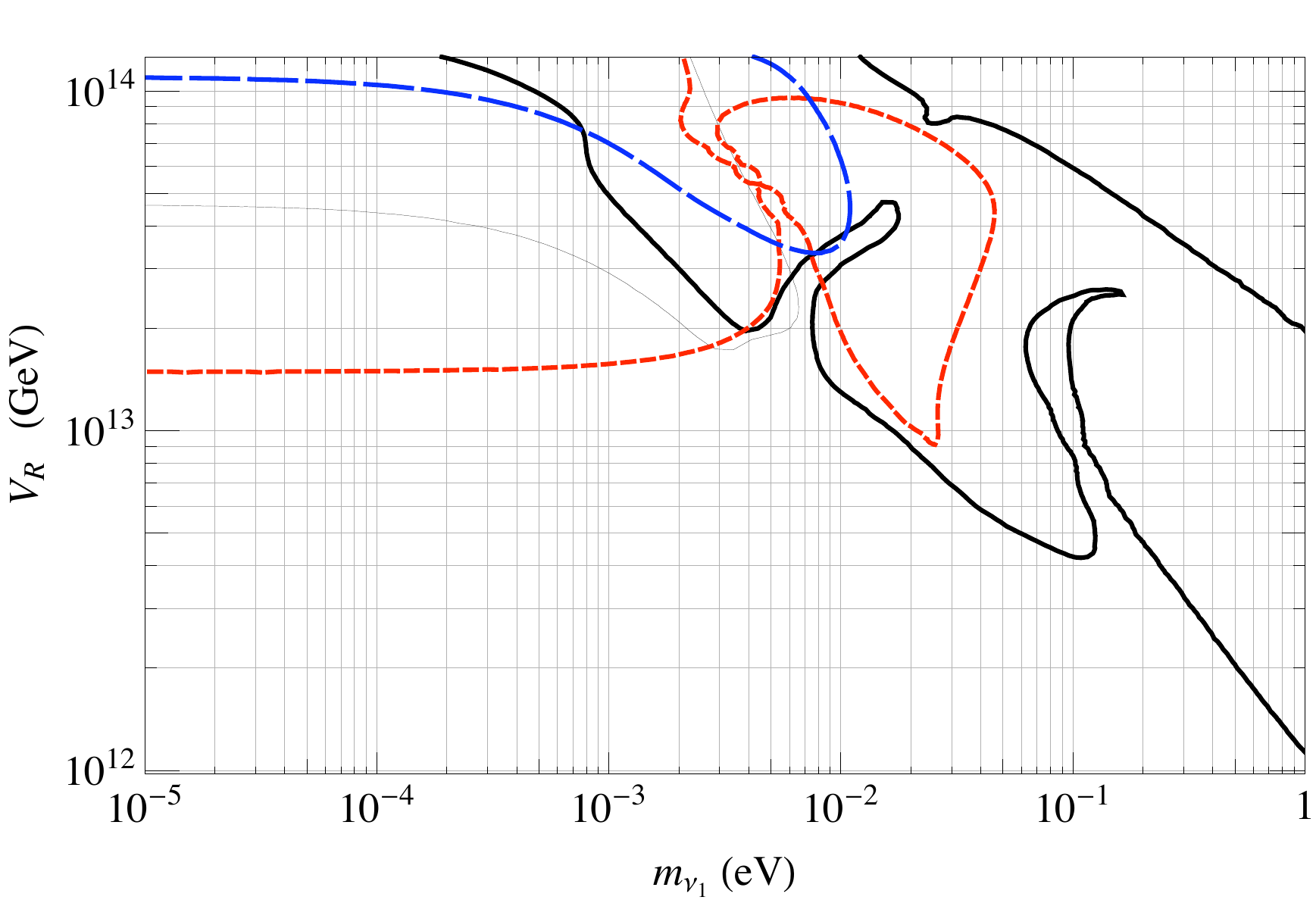} 
\caption{Regions of the ($m_1$, $v_R$) parameter space where
$|Y_B| > Y^{WMAP}_B$ for solutions $(+,+,+)$, $(+,-,+)$ and $(+,+,-)$,
and where $|Y_B| > 0.1\, Y^{WMAP}_B$ for solution $(-,-,-)$. These
regions are delimited by the thick black contour in the $(+,+,+)$ case,
the dashed red contour for $(+ ,- ,+)$, the long-dashed blue contour
for $(+,+,-)$, and the thin black contour for $(-,-,-)$. Inputs: set 1
of the Appendix for $U_m$ and the high-energy phases; other
input parameters as in Fig.~\ref{fig:Um=1}.}
\label{fig:m1_NH}
\end{center}
\end{figure}

Fig.~\ref{fig:m1_NH} shows the region of the ($m_1$, $v_R$) parameter
space where $|Y_B| > Y^{WMAP}_B$ for solutions $(+,+,+)$, $(+,-,+)$
and $(+,+,-)$, and where $|Y_B| > 0.1\, Y^{WMAP}_B$ for solution $(-,-,-)$,
which fails to generate the observed baryon asymmetry.
The choice for $U_m$ and the high-energy phases corresponds to the
set 1 described in the Appendix, the other input parameters being fixed
as in Fig.~\ref{fig:Um=1}.
In the $(+,+,+)$ case, increasing $m_1$ amounts to shift the range
of $v_R$ for which leptogenesis is successful towards lower values.
This is consistent with the fact that $M_1$ and $M_2$ grow with $m_1$,
and that the thermal production of $N_i$ is Boltzmann-suppressed
for $M_i > T_{in}$. The behaviour of the other solutions is more interesting:
in all three cases, the final baryon asymmetry is suppressed for a
quasi-degenerate light neutrino mass spectrum.
In the two solutions in which $N_2$ leptogenesis
plays a crucial role, namely $(+,+,-)$ and $(-,-,-)$, this is due to the fact that the
$N_1$-induced washout becomes strong for all flavours, as a result
of the decrease of $M_1$ (also, for $(-,-,-)$, $M_2$ decreases with growing
$m_1$). In the $(+,-,+)$ case, a larger $m_1$ implies a smaller $M_1$
and thus reduces the final baryon asymmetry.
As a result, leptogenesis fails for a quasi-degenerate light neutrino mass spectrum
in all reference solutions but $(+,+,+)$.
For the input parameters used in Fig.~\ref{fig:m1_NH}, successful leptogenesis
requires $m_1 \lesssim 0.01$ eV for solution $(+,+,-)$, and
$m_1 \lesssim 0.05$ eV for solution $(+,-,+)$.

\subsubsection{$\theta_{13}$ and $\delta_{PMNS}$}

\begin{figure}[p]
\begin{center}
\includegraphics*[width=14cm]{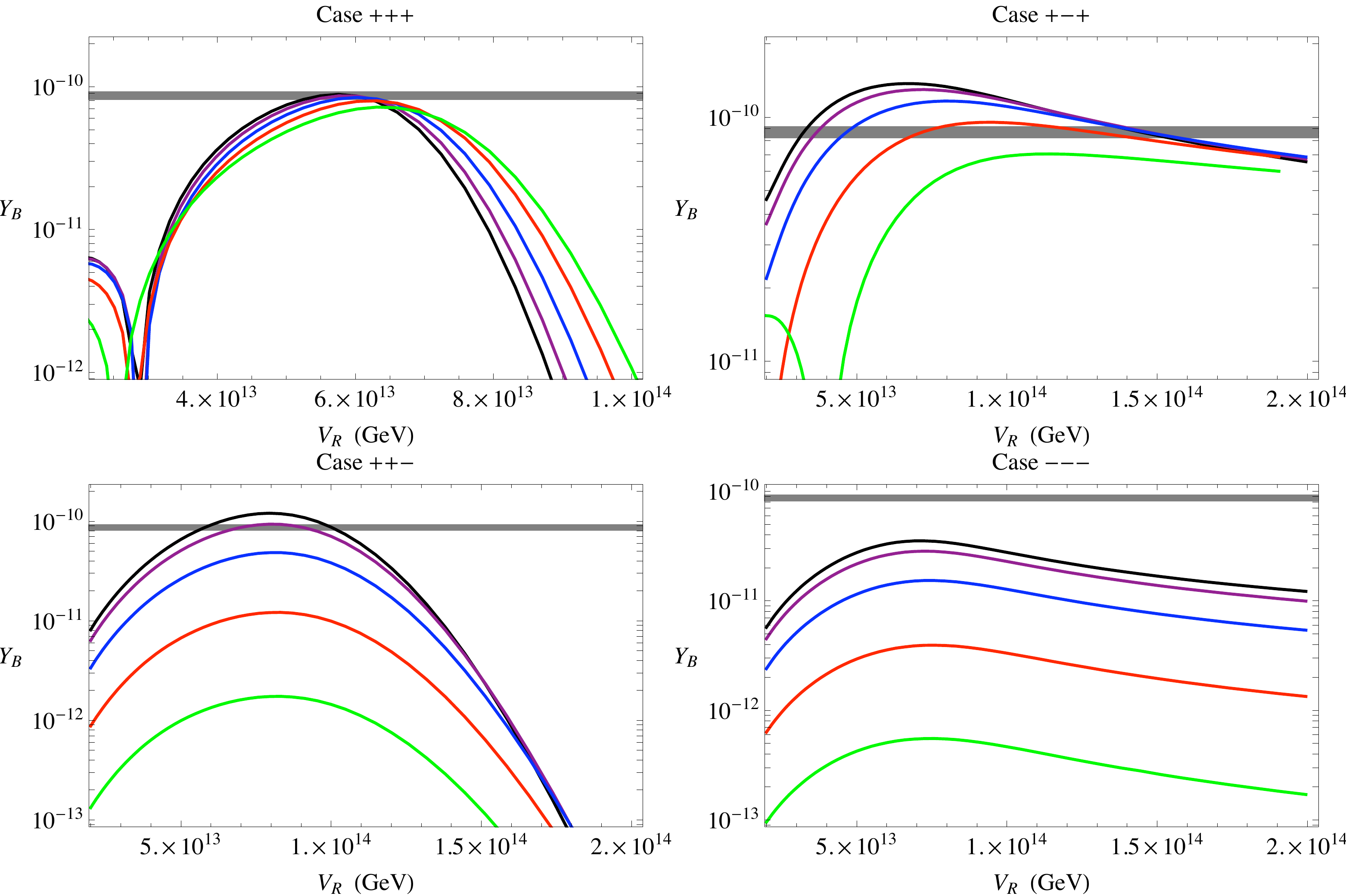}
\caption{The final baryon asymmetry as a function of $v_R$
in the four reference solutions, for $\delta_{PMNS}=0$ and different values
of $\theta_{13}$: $\theta_{13} = 0^{\circ}$ (black), $2^{\circ}$ (purple),
$5^{\circ}$ (blue), $9^{\circ}$ (red) and $13^{\circ}$ (green / light grey).
Inputs: set 1 of the Appendix for $U_m$ and the high-energy phases;
$T_{in} = 7 \times 10^9$ GeV for $(+,+,+)$ and $(+,-,+)$, while
$T_{in} = 5 \times 10^{10}$ GeV for $(+,+,-)$ and $(-,-,-)$;
other input parameters as in Fig.~\ref{fig:Um=1}.}
\label{fig:theta13}
\end{center}
\begin{center}
\includegraphics*[width=16cm]{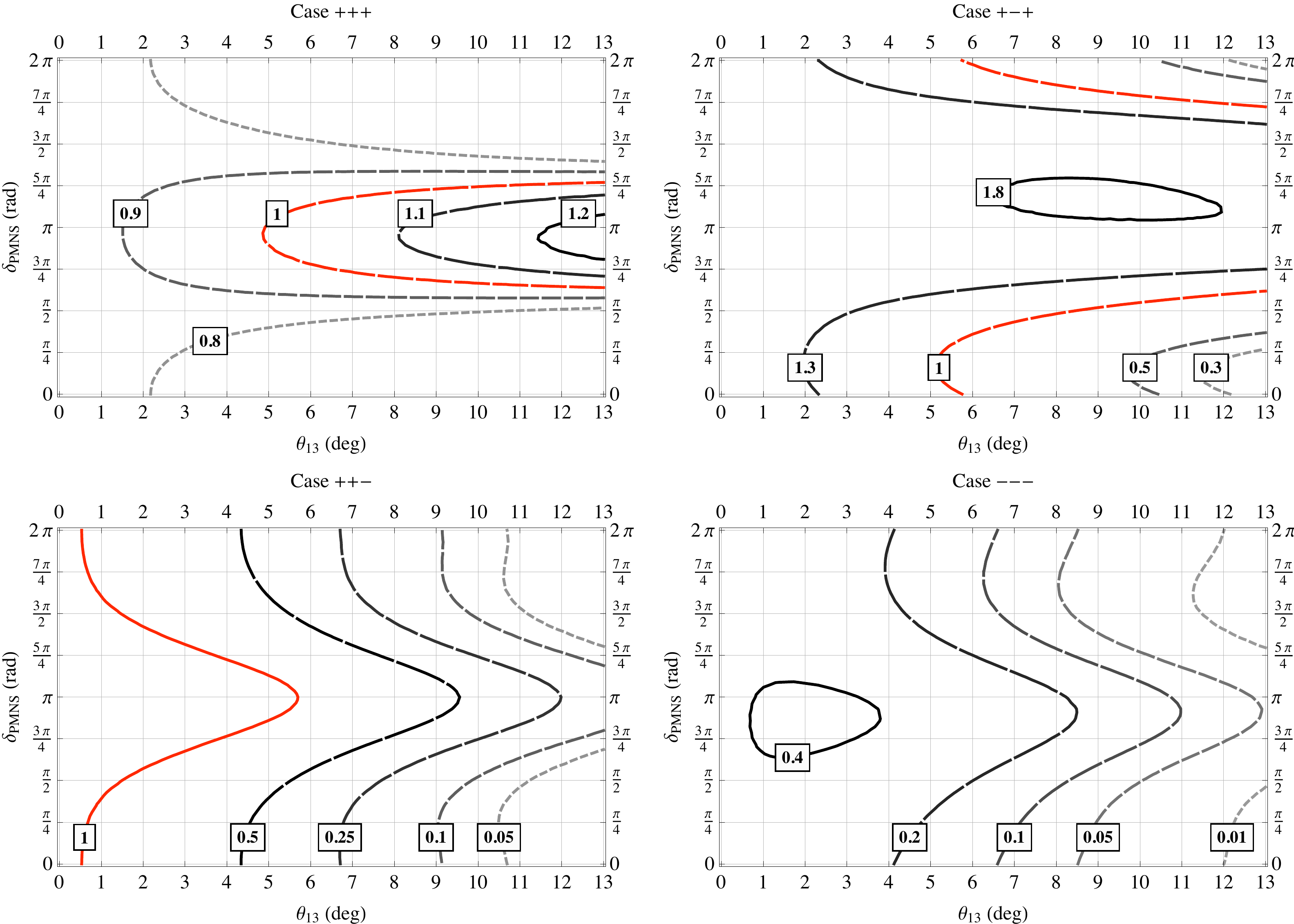}
\caption{Contour lines of the ratio $|Y_{B}| / Y_{B}^{WMAP}$ in the four
reference solutions, as a function of $\theta_{13}$ and $\delta_{PMNS}$.
The input parameters are the same as in Fig.~\ref{fig:theta13}, and the
$B-L$ breaking scale has been fixed at $v_R = 5 \times 10^{13}$~GeV
for $(+,+,+)$ and $(+,-,+)$, and at $v_R = 6 \times 10^{13}$ GeV
for $(+,+,-)$ and $(-,-,-)$.}
\label{fig:theta13_delta}
\end{center}
\end{figure}

We now turn to the dependence of the final baryon asymmetry on
$\theta_{13}$ and $\delta_{PMNS}$, the two unknown light neutrino
parameters which control the amount of CP violation in oscillations.
In Fig.~\ref{fig:theta13}, we first show the effect of varying $\theta_{13}$
alone, from our reference value $\theta_{13} = 0^\circ$ to the experimental
upper limit $\theta_{13} = 13^\circ$, assuming $\delta_{PMNS} = 0$.
The choice for $U_m$ and the high-energy phases corresponds to the
set 1 of the Appendix. Furthermore, the Boltzmann equations
are evolved from a somewhat lower initial temperature than in the previous
plots: $T_{in} = 7 \times 10^9$ GeV for $(+,+,+)$ and $(+,-,+)$, and
$T_{in} = 5 \times 10^{10}$ GeV for $(+,+,-)$ and $(-,-,-)$. The other
input parameters are chosen as in Fig.~\ref{fig:Um=1}. As can be seen
from Fig.~\ref{fig:theta13}, increasing $\theta_{13}$ generally reduces
the final baryon asymmetry, especially in solutions $(\pm,\pm,-)$
where the effect is particularly pronounced. Solution $(+,+,+)$ behaves
differently, although in this case too the maximum value of $Y_B$
is obtained for small values of $\theta_{13}$.

One could be tempted to conclude from Fig.~\ref{fig:theta13} that
(at least for the chosen input parameters) successful leptogenesis
favours small values of $\theta_{13}$. However, it is not legitimate
to impose $\delta_{PMNS} = 0$: since $\theta_{13}$ and $\delta_{PMNS}$
always appear in combination in $U_{PMNS}$, one should study
their joint effect on the final baryon asymmetry.
This is done in Fig.~\ref{fig:theta13_delta}, for the same
choice of input parameters as in Fig.~\ref{fig:theta13}, but for a fixed
value of the $B-L$ breaking scale $v_R$. One can see
that successful leptogenesis is compatible with a ``large'' value
of $\theta_{13}$ ($\theta_{13} \gtrsim 5^\circ$) as soon as $\delta_{PMNS}$
is allowed to be different from zero. Such values of $\theta_{13}$ are
within the reach of upcoming reactor and first generation superbeam
experiments (for a brief review, see e.g. Ref.~\cite{Schwetz07}).
For instance, $\theta_{13}$ just below
the present experimental limit together with $\delta_{PMNS} \approx 5 \pi / 8$
is compatible with $Y_B = Y^{WMAP}_B$ both in the $(+,+,+)$
and in the $(+,-,+)$ solution.
Although Fig.~\ref{fig:theta13_delta} was obtained for a specific set of
input parameters, we can conclude that successful leptogenesis
is possible for values of $\theta_{13}$ and $\delta_{PMNS}$
such that CP violation in the leptonic sector can be established
in future neutrino superbeam experiments.

\subsubsection{Inverted mass hierarchy}

\begin{figure}
\begin{center}
\includegraphics*[scale=0.45]{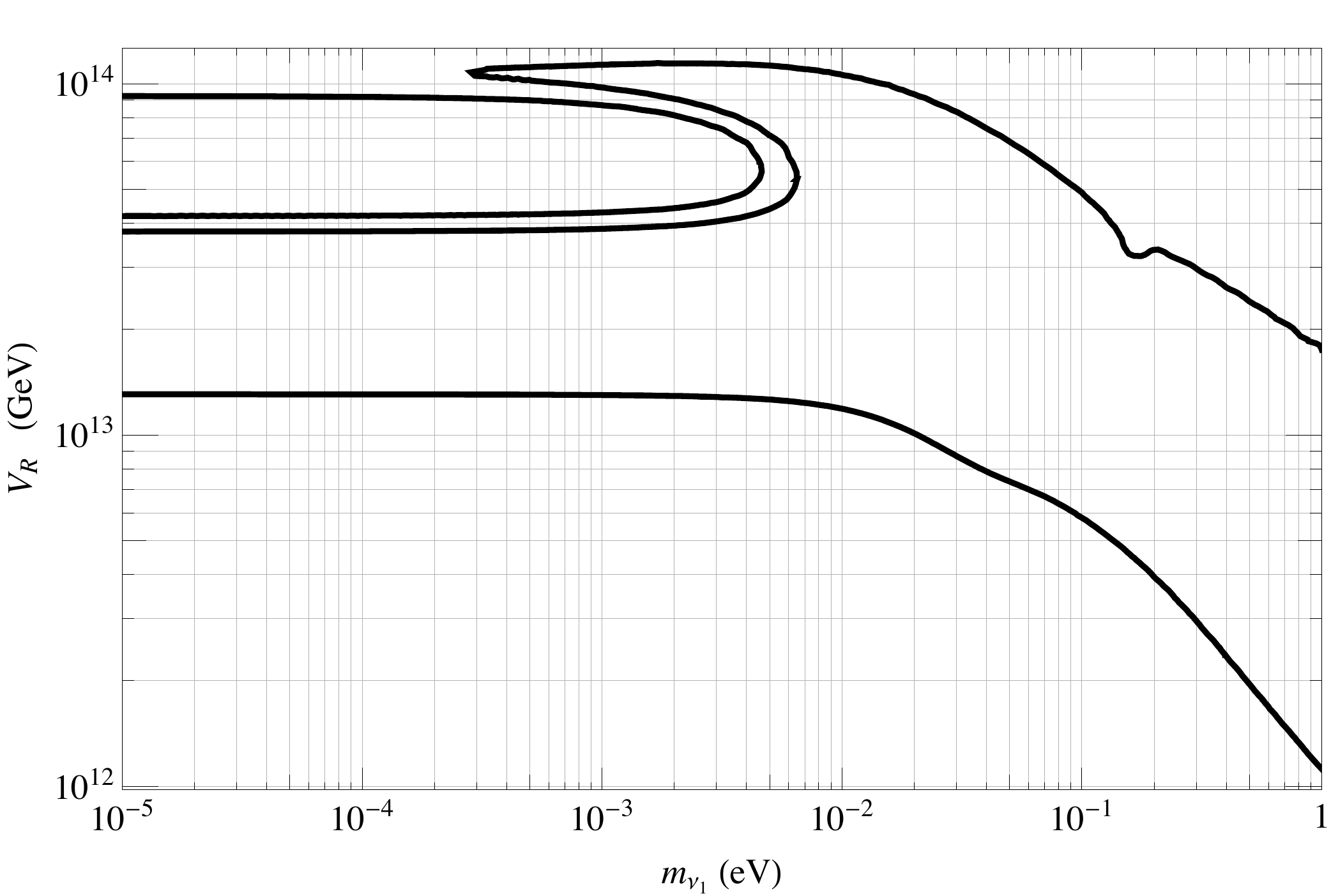} 
\caption{Region of the ($m_3$, $v_R$) parameter space where
$|Y_B| > Y^{WMAP}_B$ in the $(+,+,+)$ solution, assuming an inverted light
neutrino mass hierarchy (delimited by the black contour). Inputs: set 1
of the Appendix for $U_m$ and the high-energy phases; other
input parameters as in Fig.~\ref{fig:Um=1}.}
\label{fig:m3_IH}
\end{center}
\end{figure}

Assuming an inverted light neutrino mass hierarchy leads to
significantly different results from the normal hierarchy case,
although the gross qualitative features are preserved (in particular,
solution $(+,+,+)$ generally leads to successful leptogenesis, while
for solution $(+,-,+)$ this depends on the choice of $U_m$ and
of the Majorana and high-energy phases). In this paper, we do not
attempt to perform a general study of the inverted hierarchy case,
which has already been investigated in Ref.~\cite{ABHKO06},
in the one-flavour approximation and assuming $M_d = M_e$.
To illustrate some of the differences with the normal hierarchy 
case, we just display in Fig.~\ref{fig:m3_IH} the plot analogous to
the one in Fig.~\ref{fig:m1_NH}. Only solution $(+,+,+)$ is represented there,
since for the choice of input parameters made in Fig.~\ref{fig:m1_NH}
none of the other reference solutions leads to successful
leptogenesis in the inverted hierarchy case. This is already
a noticeable difference with the normal hierarchy case.
As far as solution $(+,+,+)$ is concerned, some differences
with respect to the normal hierarchy case can be seen
in the region where the lightest neutrino mass lies below a few
$10^{-3}$ eV. We were not able to reproduce the result
of Ref.~\cite{ABHKO06}, which finds that solution $(+,-,+)$
can generate the observed baryon asymmetry in the absence
of corrections to the mass relation $M_d = M_e$,
in the region where $M_1 \approx M_2$.
This may be due to the differences
in the treatment of leptogenesis: the study of Ref.~\cite{ABHKO06} 
was performed in the one-flavour approximation, assuming initial
thermal abundances for the right-handed neutrinos, while we solved
the flavour-dependent Boltzmann equations with $T_{in} = 10^{11}$ GeV
and vanishing initial abundances. Furthermore, Ref.~\cite{ABHKO06}
used the non-supersymmetric analogue of Eq.~(\ref{eq:f_I}) in the
computation of the CP asymmetry, while we used Eq.~(\ref{eq:epsilon_I_deg}),
which reproduces the correct behaviour of the self-energy contribution
in the limit of exactly degenerate right-handed neutrinos.

\subsection{Impact of corrections to the mass relation $M_D = M_u$}

In the preceding subsections, we studied the dependence of the final
baryon asymmetry on the values of the yet unmeasured light neutrino
parameters. Let us now turn to the influence of the high-energy
Dirac couplings.
So far we assumed that the mass relation $M_D = M_u$ holds
at the GUT scale, while $M_d = M_e$ receives corrections from
non-renormalizable operators. In this subsection, we study the effect
of a departure from $M_D = M_u$.
More specifically, we 
assume that $M_D$ and $M_u$ are still diagonal in the same
basis\footnote{This is a natural assumption if the CKM matrix mainly
comes from the down quark sector. In this case, and
in the absence of cancellations between the different contributions
to $M_D$ and $M_u$, both matrices have a strong hierarchical structure
with mixing angles smaller than the CKM angles. The relative rotation
between the bases in which $M_D$ and $M_u$ are diagonal can then
be neglected in the reconstruction procedure.}
but that their eigenvalues differ
($y_i \neq y_{u_i}$). This has a direct impact on the right-handed
neutrino mass spectrum, since the $M_i$ associated with some $x^-_j$
is to a good approximation proportional to $y^2_j$ in the regime
$v_R \gg 2 \sigma_u v^4_u / M_\Delta |z_j|^2$, while the $M_i$
associated with some $x^+_j$ is independent of $y_j$ 
(see the Appendix B of Ref.~\cite{HLS06}).
In particular, one has $M_1 \propto y^2_2$ in solution $(+,-,+)$ and
$M_2 \propto y^2_2$ in solution $(-,-,-)$. One thus expects
that raising $y_2$ will enhance the final baryon asymmetry
by increasing the $\epsilon_{1\alpha}$'s in the former case,
and the $\epsilon_{2\alpha}$'s in the latter case.

\begin{figure}
\begin{center}
\includegraphics*[height=5.5cm]{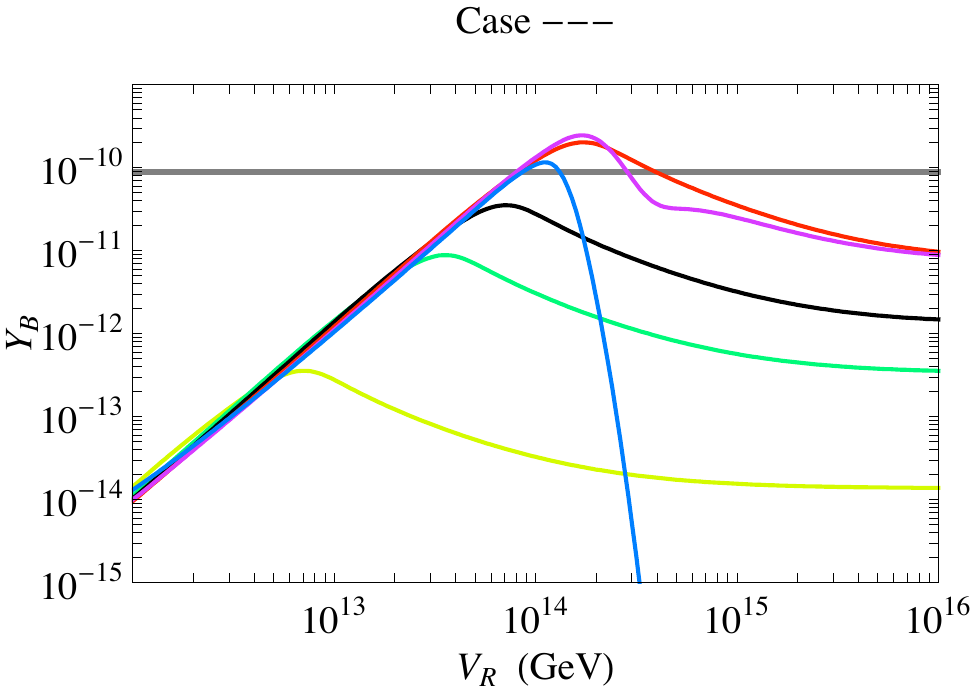} \hskip 0.5cm
\includegraphics*[height=5.5cm]{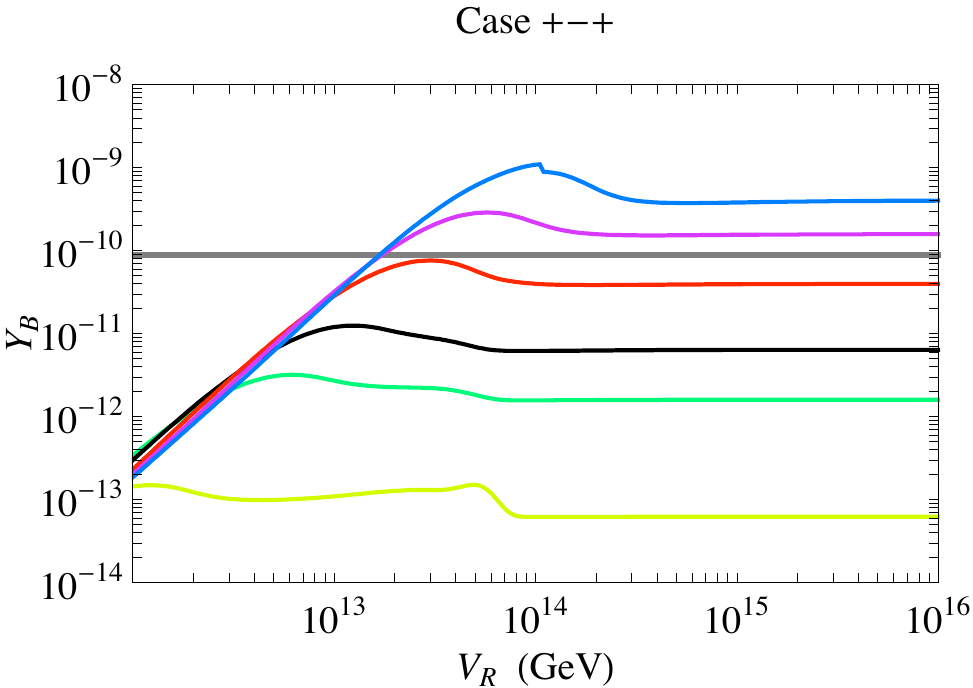} \\
\caption{The final baryon asymmetry as a function of $v_R$ for different
values of $y_2$, from $y_2 / y_c (M_{GUT}) = 0.1$ (yellow/light grey) to 
$y_2 / y_c (M_{GUT}) = 10$ (blue/dark grey). The reference case $y_2 = y_c$
is plotted in black. Left panel: solution $(-,-,-)$, set 1 for $U_m$ and the high-energy
phases; right panel: solution $(+,-,+)$, set 4 for $U_m$ and the high-energy
phases. The other input parameters are as in Fig.~\ref{fig:Um=1}.}
\label{fig:y2}
\end{center}
\end{figure}

This is shown in Fig.~\ref{fig:y2}, in which $(y_2 / y_c) (M_{GUT})$
is varied between $0.1$ and $10$ in solutions $(+,-,+)$ (right panel)
and $(-,-,-)$ (left panel). We can see that the final baryon
asymmetry increases with growing $y_2$ in both solutions. In particular,
successful leptogenesis becomes possible in the $(-,-,-)$ case for large
enough $y_2$ (for $y_2 = 10\, y_c$, however, $N_2$ becomes
too heavy to be thermally produced above $v_R \sim 10^{14}$ GeV, which
results in the Boltzmann suppression of $Y_B$). This conclusion is however
dependent on the input $T_{in} = 10^{11}$ GeV: it does not hold for
the more realistic choice $T_{in} = 10^{10}$ GeV (see the discussion
in the next subsection about the gravitino problem).
In the $(+,-,+)$ case, successful leptogenesis is possible for values
of $v_R$ as large as a few $10^{16}$ GeV, and this conclusion also
holds for $T_{in} = 10^{10}$ GeV.
This is an interesting result, since gauge coupling unification favours
a one-step breaking of the $SO(10)$ symmetry, with a $B-L$ breaking scale
close to the GUT scale (a lower $B-L$ breaking scale is however not
excluded~\cite{ABMRS00}).
Fig.~\ref{fig:y2} also shows that $y_2 > y_c$ allows
solution $(+,-,+)$ to be successful with a $U_m$ containing
only small mixing angles (set 4), thus alleviating the constraints on
the superpartner spectrum coming from the non-observation
of lepton flavour violating processes such as $\mu \rightarrow e \gamma$.

\subsection{Dependence on the reheating temperature}

\begin{figure}
\begin{center}
\includegraphics[scale=0.5]{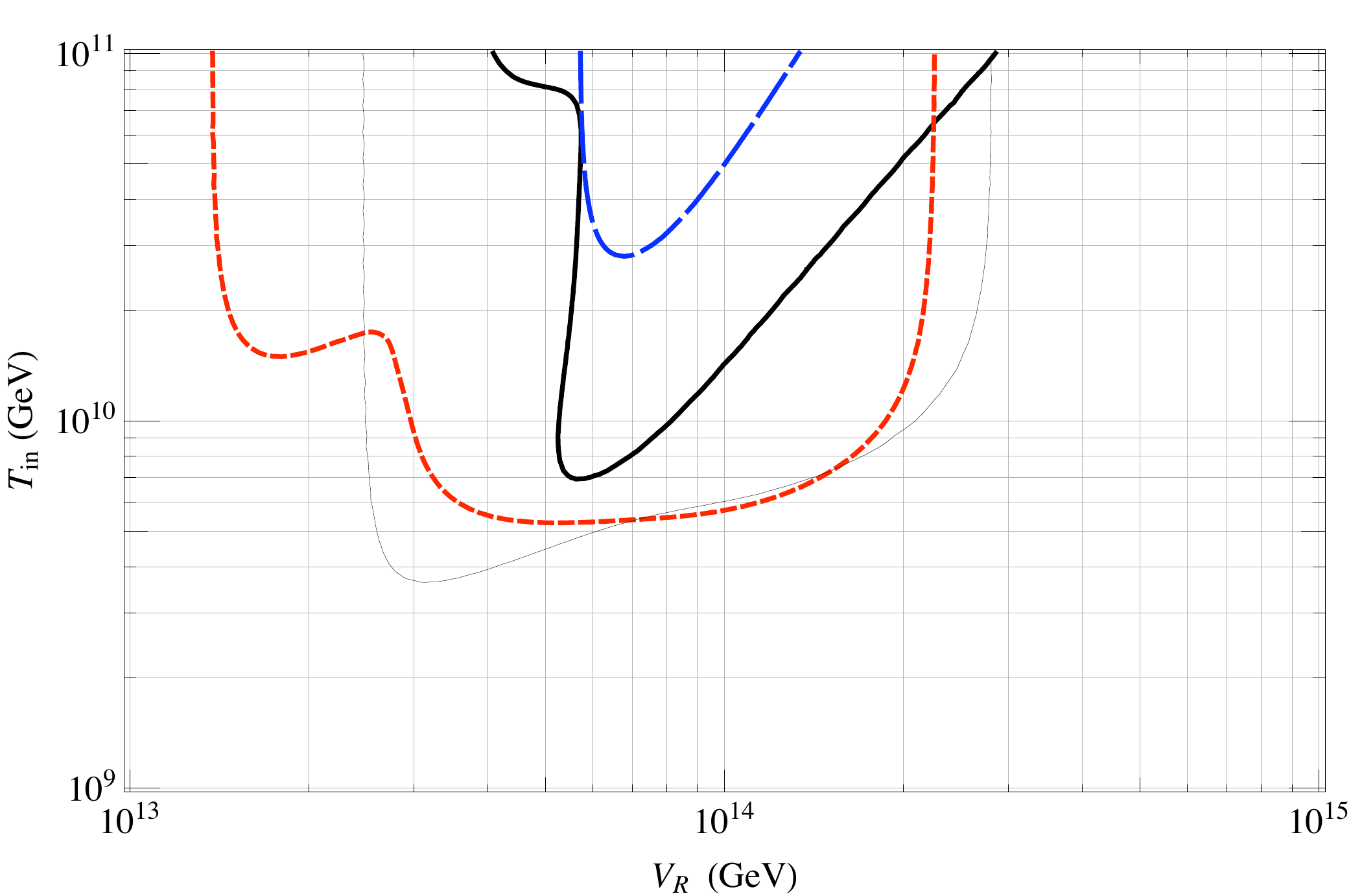}
\caption{Regions of the ($v_R$, $T_{in}$) parameter space where
$|Y_B| > Y^{WMAP}_B$ for solutions $(+,+,+)$, $(+,-,+)$ and $(+,+,-)$,
and where $|Y_B| > 0.1\, Y^{WMAP}_B$ for solution $(-,-,-)$. These
regions are delimited by the thick black contour in the $(+,+,+)$ case,
the dashed red contour for $(+ ,- ,+)$, the long-dashed blue contour
for $(+,+,-)$, and the thin black contour for $(-,-,-)$. Inputs: set 1
of the Appendix for $U_m$ and the high-energy phases; other
input parameters as in Fig.~\ref{fig:Um=1}.}
\label{fig:T_RH}
\end{center}
\end{figure}

The numerical results presented so far were obtained starting
the evolution of the Boltzmann equations at $T_{in} = 10^{11}$ GeV,
in the approximation where the dynamics of reheating is neglected.
In this approach, $T_{in}$ can be identified with the reheating temperature.
In order to estimate how severe the tension between successful leptogenesis
and the gravitino problem is, we therefore proceed to study
the dependence of the final baryon asymmetry on $T_{in}$.
Fig.~\ref{fig:T_RH} shows the regions of the ($v_R$, $T_{in}$)
parameter space where $|Y_B| > Y^{WMAP}_B$ for solutions $(+,+,+)$,
$(+,-,+)$ and $(+,+,-)$, and where $|Y_B| > 0.1\, Y^{WMAP}_B$
for solution $(-,-,-)$.
The choice for $U_m$ and the high-energy phases corresponds to the
set 1 of the Appendix, the other input parameters being fixed as in
Fig.~\ref{fig:Um=1}.
One can see that solution $(+,-,+)$
succeeds in generating the observed baryon asymmetry for values
of $T_{in}$ as low as $5 \times 10^9$ GeV, whereas solutions $(+,+,+)$
and $(+,+,-)$ require $T_{in} \gtrsim 7 \times 10^9$ GeV and
$T_{in} \gtrsim 3 \times 10^{10}$ GeV, respectively. While these
numbers have been obtained for a particular choice of the input
parameters,
they unambiguously show that successful leptogenesis can
be achieved with a reheating temperature below $10^{10}$ GeV
in solutions $(+,+,+)$ and $(+,-,+)$. As for solution $(+,+,-)$,
$T_{in} > 10^{10}$ GeV was found to be a necessary condition
for successful leptogenesis for all sets of input parameters we considered.
This allows us to conclude that, for generic input parameters,
the solution $(+,+,-)$ fails to generate the observed baryon asymmetry
if the reheating temperature is lower than $10^{10}$ GeV.

As discussed at the end of Section~\ref{sec:leptogenesis},
there are strong constraints on the reheating temperature
from gravitino cosmology, and this potentially conflicts with
successful thermal leptogenesis. Nevertheless, some supersymmetric
scenarios can accommodate a reheating temperature in the
$(10^9 - 10^{10})$ GeV range, as required for solutions $(+,+,+)$
and $(+,-,+)$ to generate the correct amount of baryon asymmetry.
One possibility is that the gravitino
is the LSP; the constraint that its relic density does not exceed
the dark matter abundance reads $T_{RH} \lesssim (10^9 - 10^{10})$ GeV
for $m_{3/2} \sim 100$ GeV~\cite{thermal_production}.
This scenario is further constrained by the requirement that
the NLSP decays do not alter the success of Big Bang nucleosynthesis (BBN),
which can be satisfied e.g. by a sneutrino NLSP~\cite{KKKM06}
or by assuming some amount of $R$-parity violation~\cite{BCHIY07}.
Another way of avoiding the strong constraints on the reheating
temperature is to assume an extremely light gravitino~\cite{PP81},
$m_{3/2} \leq 16$ eV~\cite{VLHMR05} (where the upper
bound comes from WMAP and Lyman-$\alpha$ forest data),
or a very heavy gravitino~\cite{Weinberg82}, $m_{3/2} \gtrsim 50$~TeV.
In the former case, the gravitino decouples when it is still relativistic
and escapes the overproduction problem, while the NLSP is sufficiently
short-lived to decay before BBN. In the latter case, the gravitino decays
before nucleosynthesis and does not affect the light element
abundances; furthermore, the LSPs produced in its decays
are within the observed dark matter abundance for
$T_{RH} \lesssim 10^{10}$ GeV~\cite{GGW99}.

\section{Conclusions}
\label{sec:conclusions}

In this paper, we studied thermal leptogenesis in a broad class of
supersymmetric $SO(10)$ models with a left-right symmetric
seesaw mechanism, including flavour effects and the contribution
of the next-to-lightest right-handed neutrino supermultiplet.
Assuming $M_D = M_u$ and a normal hierarchy of light neutrino
masses, we found that successful
leptogenesis is possible with a reheating temperature in the
($10^9 - 10^{10}$) GeV range for 4 out of the 8 reconstructed
right-handed neutrino mass spectra\footnote{Although we only presented
results for four reference solutions, we checked that the solutions
that differ by $x^+_1 \leftrightarrow x^-_1$, such as $(+,-,+)$
and $(-,-,+)$, show similar behaviour for $Y_B$.},
corresponding to solutions $(\pm, +, +)$ and $(\pm,-,+)$.
In the remaining 4 solutions, leptogenesis is dominated by $N_2$
decays, as in the type I seesaw case; among those, solutions
 $(\pm,+,-)$ succeed in generating the observed baryon asymmetry for
 reheating temperatures above $10^{10}$ GeV.
These results show that $SO(10)$ models in which
the Dirac mass matrix and the up quark mass matrix
have a similar hierarchical structure
are compatible with successful thermal leptogenesis
if the seesaw mechanism is of the left-right symmetric type.
As a byproduct, we found that solution $(-,-,-)$, which mimics
the type I seesaw in the large $v_R$ limit, fails to
generate the right amount of baryon asymmetry in this limit.
This suggests that successful flavour-dependent $N_2$ leptogenesis
in $SO(10)$ models with a type I seesaw mechanism
requires large corrections to the mass formula $M_D = M_u$,
especially if one insists on $T_{RH} < 10^{10}$ GeV.

Both flavour effects and the corrections to the mass
relation $M_d = M_e$ were found to be crucial ingredients for the success
of solutions $(\pm,-,+)$ and, for $T_{RH} > 10^{10}$ GeV,
of solutions $(\pm,+,-)$. In the former case, flavour effects increase
the final baryon asymmetry by up to one order of magnitude,
while in the latter case they allow a particular lepton flavour asymmetry
generated in $N_2$ decays to survive
the washout by $N_1$-related processes, resulting in a spectacular
enhancement of the final baryon asymmetry with respect to
the one-flavour approximation.
We also studied the dependence of the results on the unknown
light neutrino parameters ($\theta_{13}$, $\delta_{PMNS}$ and
the type of mass hierarchy), as well as on corrections to the
mass relation $M_D = M_u$ and on the reheating temperature.
Moderate deviations from $M_D = M_u$ were shown to extend
the region of parameter space in which leptogenesis is successful;
in particular, solution $(+,-,+)$ was found to be successful for
values of the $B-L$ breaking scale as large as $10^{16}$~GeV,
as preferred by supersymmetric gauge coupling unification.

We believe that the results presented in this paper (which extend
and put on a more solid basis the ones of Refs.~\cite{HLS06,ABHKO06})
make $SO(10)$ models in which neutrino masses originate from
the left-right symmetric seesaw mechanism more attractive.
Other aspects of the phenomenology of these models,
such as lepton flavour violation (which was briefly discussed
in Ref.~\cite{HLS06}), could discriminate further between
the 8 different right-handed neutrino mass spectra,
and are worth studying in detail.
We note in particular that the typical values of the mixing angles
in $U_m$ allowing solution $(+,-,+)$ to be successful tend
to enhance the branching ratios of processes like $\mu \rightarrow
e \gamma$ and $\tau \rightarrow \mu \gamma$.

Finally, it would be interesting to investigate how the results presented
in this paper would be modified if the $SU(2)_L$ doublet components
of the $\bf \overline{126}$ were allowed to acquire a vev. In such a case,
the mass relations $M_d = M_e$ and $M_D = M_u$
receive corrections proportional to the matrix $f$,
and the measured down quark and charged lepton masses can
be accounted for without introducing non-renormalizable interactions.
However, a technical complication arises from the fact that the reconstruction
procedure cannot be performed independently of the charged
fermion mass fit. We defer the study of this case to future work.


\section*{Acknowledgments}
We thank M.~Frigerio, T.~Hambye and J.-M.~Fr\`ere for useful comments,
as well as T.~Konstandin for discussions about Ref.~\cite{ABHKO06}.
This work has been supported in part by the projects ANR-05-BLAN-0193-02
and ANR-05-JCJC-0023 of the French Agence  Nationale de la Recherche.
PH and SL acknowledge partial support from the RTN European Program
MRTN-CT-2004-503369. The research of PH was supported in part by
the Marie Curie Excellence Grant MEXT-CT-2004-01429717.


\renewcommand{\theequation}{A.\arabic{equation}}
\setcounter{equation}{0}  

\begin{appendix}

\section{Appendix: corrections to the mass relation $M_d = M_e$}
\label{app:corrections}

In this appendix, we discuss the corrections to the GUT-scale
mass relation $M_d = M_e$ arising from non-renormalizable operators
of the form:
\beq
\frac{\kappa_{ij}}{\Lambda}\ {\bf 16}_i {\bf 16}_j
  \left( {\bf 10}_1 \times {\bf 45} \right)_{| \bf 120}\ ,
\label{eq:nr_ops}
\eeq
where only the $Y\! =\! -1$  $SU(2)_L$ doublet in ${\bf 10}_1$
acquires a vev, so that the mass relation $M_D = M_u$ is not affected,
and the ${\bf 10}_1$ and  $\bf 45$ Higgs representations are contracted
in an effective $\bf 120$ representation,
implying that the couplings $\kappa_{ij}$ are antisymmetric.
Such operators generate a contribution to the down quark
and charged lepton masses when the $\bf 45$
acquires a vev in the $T_{3R}$ or $B-L$ direction.
Indeed, the decomposition of the tensor product $\mathbf{(10 \times 45)_{|120}}$
under the Pati-Salam subgroup $SU(2)_L\times SU(2)_R\times SU(4)_c$
contains a $\mathbf{(2,2,1)}$ representation generated by
$\mathbf{(2,2,1)_{10}\times(1,3,1)_{45}}$,
and a $\mathbf{(2,2,15)}$ representation generated by
$\mathbf{(2,2,1)_{10}\times(1,1,15)_{45}}$. The Clebsch-Gordan coefficients
needed to distinguish $M_d$ from $M_e$ arise from $\mathbf{(2,2,15)}$.

The most general situation occurs when the two vev's $\mathbf{\md{(1,3,1)}}$
in the $T_{3R}$ direction and $\mathbf{\md{(1,1,15)}}$ in the $B-L$ direction
belong to different $\mathbf{45}$'s. We denote the scale of these vev's by
$v_3$ and $v_{15}$ and assume that they have GUT scale values.
The other dimensionful parameter appearing in Eq. (\ref{eq:nr_ops})
is $\Lambda$, which we identify with the scale at which the unified
gauge coupling becomes non-perturbative,
$\Lambda \simeq 10\, M_{GUT}$~\cite{HLS06}.
The corrected mass matrices read:
\begin{equation}
M_d\ =\ M^{10}_d + \left( -\frac{v_3}{\Lambda}\, \kappa_1
  +\frac{v_{15}}{\Lambda}\, \kappa_2 \right) v^{10_1}_d\ , \qquad
M_e\ =\ M^{10}_d + \left( -\frac{v_3}{\Lambda}\, \kappa_1
  -3\, \frac{v_{15}}{\Lambda}\, \kappa_2 \right) v^{10_1}_d\ ,
\end{equation}
where $M^{10}_d$ is the contribution of the $\bf 10$'s,
$v^{10_1}_d$ is the vev of the $Y\! =\! -1$ Higgs doublet in $\bf 10_1$,
and the matrices
$\kappa_1$ and $\kappa_2$ contain the non-renormalizable
couplings associated with the two $\bf 45$'s.

In order to study the corrected mass matrices, it is convenient
to switch to the basis where the symmetric contribution $M_d^{10}$
is diagonal. Then $M_d$ and $M_e$ can be parametrized as:
\begin{equation}
M_d\ =\ \left(
\begin{array}{ccc}
\mu_1 & \varepsilon_1 & \varepsilon_2 \\
-\varepsilon_1 & \mu_2 & \varepsilon_3 \\
-\varepsilon_2 & -\varepsilon_3 & \mu_3 \\
\end{array} \right)\ , \qquad \qquad
M_e\ =\ \left(
\begin{array}{ccc}
\mu_1 & -x_1\varepsilon_1 & -x_2\varepsilon_2 \\
x_1\varepsilon_1 & \mu_2 & -x_3\varepsilon_3 \\
x_2\varepsilon_2 & x_3\varepsilon_3 & \mu_3
\end{array}
\right)\ ,
\label{eq:Md_Me_corrected}
\end{equation}
with $\mu_i$ real and $\varepsilon_i$, $x_i$ complex
($\mu_i$ and $\varepsilon_i$ are dimensionful parameters).
This structure simplifies in some cases. If the vev's
in the $T_{3R}$ and $B-L$ directions are carried by the same
$\mathbf{45}$, there is a single matrix of non-renormalizable
 couplings $\kappa$ and all $x_i$ are equal.
In the absence of a vev in the $T_{3R}$ direction, one has
$x_1=x_2=x_3=3$.
The matrices $M_d^\dagger M_d$ and $M_e^\dagger M_e$
can easily be diagonalized in the case of hierarchical entries,
$\mu_1\ll \mu_2\ll \mu_3$ and
$|\varepsilon_1|\ll|\varepsilon_2|\ll|\varepsilon_3|$, yielding:
\begin{eqnarray}
m_b^2 &\simeq& \mu_3^2+2 |\varepsilon_3|^2\, , \label{eq:mb2} \\
m_\tau^2 &\simeq& \mu_3^2+2 |x_3 \varepsilon_3|^2\, , \\
m_bm_s &\simeq& \left| \mu_2 \mu_3+\varepsilon_3^2 \right| , \\
m_\tau m_\mu &\simeq& \left| \mu_2 \mu_3+x_3^2\varepsilon_3^2 \right| . \label{eq:mtmm}
\end{eqnarray}
The fact that $\varepsilon^2_3$ appears in all four
relations~(\ref{eq:mb2})--(\ref{eq:mtmm}) implies that antisymmetric
corrections can at most accommodate a GUT-scale $\tau - b$
mass difference of the order of $m_\mu$, and this conclusion
also holds for more general hierarchies of the $\mu_i$ and $\varepsilon_i$
parameters. This is not enough to account for the value of $(m_\tau - m_b)$
obtained by running the measured down quark and charged lepton masses
from $M_Z$ to $M_{GUT}=2\times 10^{16}$ GeV (assuming an effective
supersymmetric threshold $M_{SUSY} = 1$ TeV and $\tan \beta=10$):
\begin{equation}
\begin{array}{ccc}
m_d\, (M_{GUT})\, =\, 0.94\, \mbox{MeV}\ ,
  & m_s\, (M_{GUT})\, =\, 17\, \mbox{MeV}\ ,
  & m_b\, (M_{GUT})\, =\, 0.98\, \mbox{GeV}\ ,  \\
m_e\, (M_{GUT})\, =\, 0.346\, \mbox{MeV}\ ,
  & m_\mu\, (M_{GUT})\, =\, 73.0\, \mbox{MeV}\ ,
  & m_\tau\, (M_{GUT})\, =\, 1.25\, \mbox{GeV}\ .
\end{array}
\label{m(MGUT)}
\end{equation}
However, we did not include in our analysis  the low-energy
supersymmetric threshold corrections to the bottom quark mass,
which can substantially modify the value of
$m_b (M_{GUT})$~\cite{Susy_thresholds}.
These corrections take the form:
\begin{equation}
m_b\ =\ (1+\epsilon_b\tan\beta)\, y_b v_d\ ,
\end{equation}
with the coefficient $\epsilon_b$ given by:
\begin{equation}
\epsilon_b\ =\ \frac{2\alpha_3}{3\pi}\, \frac{\mu M_3}{m_{\tilde{b}_R}^2}\,
  f(M_3^2,m_{\tilde{b}_L}^2,m_{\tilde{b}_R}^2)
  + \frac{y_t^2}{16\pi^2}\, \frac{\mu A_t}{m_{\tilde{b}_R}}\,
  f(\mu^2,m_{\tilde{t}_L}^2,m_{\tilde{t}_R}^2)\ ,
\end{equation}
where $f$ is a loop function defined by:
\begin{equation}
f(m_1^2,m_2^2,m_3^2)\ =\ \left[\frac{m_1^2}{m_3^2-m_1^2}\mbox{ln}\left(\frac{m_1^2}{m_3^2}\right)-\frac{m_2^2}{m_3^2-m_2^2}\mbox{ln}\left(\frac{m_2^2}{m_3^2}\right)\right]\frac{m_3^2}{m_1^2-m_2^2}\ .
\end{equation}
The function $f$ is of order one for superpartner masses between
$100$ GeV and $1$ TeV, yielding typical values of $\epsilon_b\sim 2\%$.
Due to the $\tan \beta$ enhancement, the threshold corrections
can thus reach the $20\%$ level for $\tan \beta = 10$. This is enough
to reach $m_b(M_{GUT})\simeq 1.17$ GeV, making it possible for the
antisymmetric corrections to the GUT-scale mass relation $M_d = M_e$
to account for the measured down quark and charged lepton masses.

Indeed, very good fits of the charged lepton and down quark masses
can be obtained once supersymmetric threshold corrections are taken
into account. Since the quality of the fit does not depend on the precise
values of the $x_i$, we restrict ourselves to the case
of a single $\mathbf{45}$ vev in the $\mathbf{(1,1,15)}$ direction, i.e.
we set $x_1=x_2=x_3=3$ in Eq.~(\ref{eq:Md_Me_corrected}).
For each set of parameters $(\mu_i, \varepsilon_i)$ providing a good fit,
$M_d$ and $M_e$ are determined as well as the mismatch matrix
$U_m$ introduced in Section~\ref{subsec:corrections}. Since only
$U_m$ is needed for the computation of the final baryon asymmetry,
we concentrate on this matrix from now on. Let us introduce the
following parametrization:
\begin{equation}
U_m\ =\ e^{i\phi_g^m}\left(
\begin{array}{ccc}
e^{i\phi_1^m} & 0 & 0\\
0 & e^{i\phi_2^m} & 0\\
0 & 0 & 1
\end{array}
\right) V(\theta_{12}^m,\theta_{13}^m,\theta_{23}^m,\delta^m) \left(
\begin{array}{ccc}
e^{i\phi_3^m} & 0 & 0\\
0 & e^{i\phi_4^m} & 0\\
0 & 0 & 1
\end{array}
\right)\ ,
\end{equation}
where $V$ is a CKM-like matrix with three real angles and a complex
phase. As Eq.~(\ref{eq:Md_Me_corrected}) contains more parameters
than needed to fit the down-type fermion masses, we fix some of them,
such as $|\varepsilon_1|$ and the phase of $\varepsilon_2$.
For $|\varepsilon_1| \ll |\varepsilon_2|$, which roughly corresponds
to $|\varepsilon_1| \lesssim 0.001$ GeV, good fits generally correspond
to values of the $\theta^m_{ij}$'s of the order of the Cabibbo angle
or smaller, for instance:
\begin{equation}
\theta_{12}^m\, \approx\, 0.3\ , \qquad \theta_{13}^m\, \approx\, 0.1\ ,
\qquad \theta_{23}^m\, \approx\, 0.35\ .
\end{equation}
For larger values of $|\varepsilon_1|$ ($|\varepsilon_1|\gtrsim 0.001$ GeV),
one can obtain larger $(1,2)$ and $(1,3)$ mixing angles, e.g.:
\begin{equation}
\theta_{12}^m\, \approx\, 1\ , \qquad \theta_{13}^m\, \approx\,  0.2\ ,
\qquad \theta_{23}^m\, \approx\,  0.2\ .
\end{equation}
In our numerical study of leptogenesis, we use different choices
for $U_m$, combined with a non-vanishing high-energy or Majorana
phase that we fix at $\pi/4$. The resulting four sets of parameters
are displayed in the table below:

\begin{center}
\begin{tabular}{|c|c|c|c|c|c|c|c|c|c|c|}
\hline set & $\theta_{12}^m$ & $\theta_{13}^m$ & $\theta_{23}$ & $\delta^m$ & $\phi_g^m$ & $\phi_1^m$ & $\phi_2^m$ & $\phi_3^m$ & $\phi_4^m$ & $\pi/4$ \\
\hline 1 & 1.07 & 0.22 & 0.21 & 5.80 & 3.21 & 4.37 & 5.86 & 0.87 & 6.16 & $\Phi_2^u$ \\
\hline 2 & 1.07 & 0.22 & 0.21 & 5.80 & 3.21 & 4.37 & 5.86 & 0.87 & 6.16 & $\Phi_2^\nu$ \\
\hline 3 & 0.28 & 0.089 & 0.37 & 0.062 & 3.15 & 3.12 & 6.03 & 2.94 & 6.19 & $\Phi_2^u$ \\
\hline 4 & 0.17 & 0.066 & 0.29 & 0.23 & 3.14 & 0.54 & 0.015 & 6.27 & 0.0032 & $\Phi^u_2$ \\
\hline
\end{tabular}
\end{center}

\begin{figure}[h!]
\begin{center}
\includegraphics[width=8.cm]{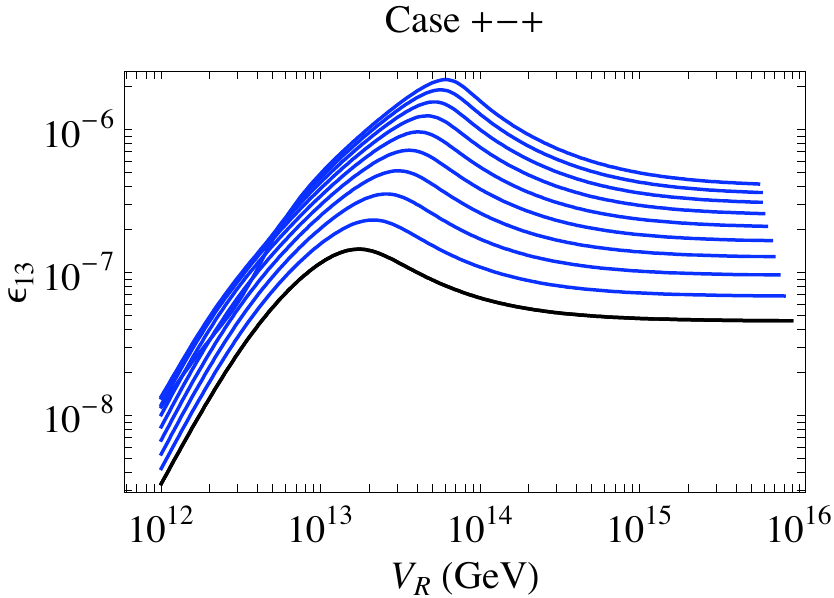} \hspace{0.5cm}
\includegraphics[width=8.cm]{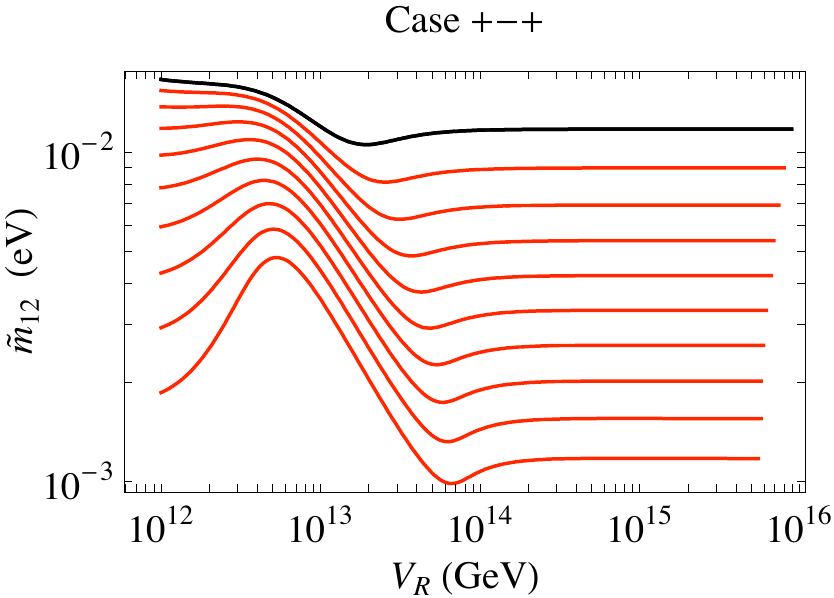}
\caption{$\varepsilon_{1\tau}$ and $\tilde{m}_{1\mu}$ as a function of $v_R$
in the $(+,-,+)$ solution, for $\Phi^u_2 = \pi/4$ and different values of
$\theta_{12}^m\in[0,\pi/4]$. All other parameters in $U_m$ and high-energy
phases are set to zero. The reference case $\theta_{12}^m = 0$ ($U_m = \unit$)
is plotted in black. The other input parameters are as in Fig.~\ref{fig:Um=1}.}
\label{fig:theta12_m}
\end{center}
\end{figure}

To illustrate the influence of $U_m$ on the leptogenesis parameters,
we plot in Fig.~\ref{fig:theta12_m} the CP asymmetry $\epsilon_{1\tau}$
and the washout parameter $\tilde m_{1\mu}$ for different values of 
$\theta^m_{12}$. Part of the effect which can be seen is due to the influence
of the $\theta_{ij}^m$'s on the right-handed neutrino masses.

\end{appendix}



\end{document}